%% file: draft_Lc2LmdX_arXiv.tex
\documentclass[prl,twocolumn,showpacs,superscriptaddress,amsmath,amssymb,nofootinbib]{revtex4-1}
\usepackage{graphicx}
\usepackage{epsfig}
\usepackage{bigstrut}
\usepackage{overpic}
\usepackage{dcolumn}
\usepackage{bm}
\usepackage{color}
\usepackage{amsmath}
\usepackage{amsfonts}
\usepackage{lineno}
\usepackage{xspace}
\usepackage{multirow}
\usepackage{subfigure}
\usepackage{verbatim}
\usepackage{footmisc}
\usepackage{epstopdf}
\usepackage{xcolor}
\usepackage{soul}
\usepackage{dcolumn}
\usepackage{float}
\RequirePackage{lineno}
\usepackage[colorlinks,linkcolor=blue,anchorcolor=blue,citecolor=blue]{hyperref}

\hypersetup{hypertex=true,
	colorlinks=true,
	linkcolor=blue,
	anchorcolor=blue,
	citecolor=blue}

\newcommand{\jpsi}{J/\psi}

\newcommand{\bfg}{\begin{figure}}
	\newcommand{\efg}{\end{figure}}
\newcommand{\bitm}{\begin{itemize}}
	\newcommand{\eitm}{\end{itemize}}
\newcommand{\bnum}{\begin{enumerate}}
	\newcommand{\enum}{\end{enumerate}}
\newcommand{\btbl}{\begin{table}}
	\newcommand{\etbl}{\end{table}}
\newcommand{\btbu}{\begin{tabular}}
	\newcommand{\etbu}{\end{tabular}}

\newcommand{\beq}{\begin{equation}}
\newcommand{\edq}{\end{equation}}

\newcommand{\ee}{e^{+}e^{-}}

\newcommand{\pip}{\pi^{+}}
\newcommand{\pim}{\pi^{-}}

\newcommand{\Lmd}{\Lambda}
\newcommand{\Lmb}{\bar{\Lambda}}

\newcommand{\Lcp}{\Lambda_{c}^{+}}
\newcommand{\Lcm}{\bar{\Lambda}_{c}^{-}}

\newcommand{\Lmdtoppim}{\Lmd \to p \pim}

\newcommand{\modea}{pK_{S}^0}
\newcommand{\modeb}{pK^{-}\pi^+}
\newcommand{\modec}{pK_{S}^0\pi^0}
\newcommand{\moded}{pK_{S}^0\pi^+\pi^-}
\newcommand{\modee}{pK^{-}\pi^+\pi^0}

\newcommand{\modeaa}{\Lambda\pi^+}
\newcommand{\modebb}{\Lambda\pi^+\pi^0}
\newcommand{\modedd}{\Lambda\pi^+\pi^-\pi^+}
\newcommand{\modeaaa}{\Sigma^{0}\pi^+}
\newcommand{\modeccc}{\Sigma^{+}\pi^0}
\newcommand{\modeddd}{\Sigma^{+}\pi^+\pi^-}

\newcommand{\Modea}{\bar{p}{}K_{S}^0}
\newcommand{\Modeb}{\bar{p}{}K^{+}\pi^-}
\newcommand{\Modec}{\bar{p}{}K_{S}^0\pi^0}
\newcommand{\Moded}{\bar{p}{}K_{S}^0\pi^-\pi^+}
\newcommand{\Modee}{\bar{p}{}K^{+}\pi^-\pi^0}

\newcommand{\Modeaa}{\bar{\Lambda}{}\pi^-}
\newcommand{\Modebb}{\bar{\Lambda}{}\pi^-\pi^0}
\newcommand{\Modedd}{\bar{\Lambda}{}\pi^-\pi^+\pi^-}
\newcommand{\Modeaaa}{\bar{\Sigma}{}^{0}\pi^-}
\newcommand{\Modeccc}{\bar{\Sigma}{}^{-}\pi^0}
\newcommand{\Modeddd}{\bar{\Sigma}{}^{-}\pi^-\pi^+}

\newcommand{\LcptoLX}{\Lcp \to \Lmd X}
\newcommand{\LcmtoLX}{\Lcm \to \Lmb X}

\def\<{\langle}
\def\>{\rangle}

\newcommand{\mBC}{M_{\rm{BC}}}

\newcommand{\gev}{\,\rm{GeV}}

\newcommand{\gevcc}{\,\rm{GeV}/\it{c}^{\rm{2}}}

\newcommand{\acpd}{\mathcal{A}_{\rm{CP}}^{\rm{dir}}}
\newcommand{\acpa}{\mathcal{A}_{\rm{CP}}^{\rm{pol}}}

\newcolumntype{d}[1]{D{.}{.}{#1}}

\usepackage{siunitx}
\newcommand{\BESIIIorcid}[1]{\href{https://orcid.org/#1}{\hspace*{0.1em}\raisebox{-0.45ex}{\includegraphics[width=1em]{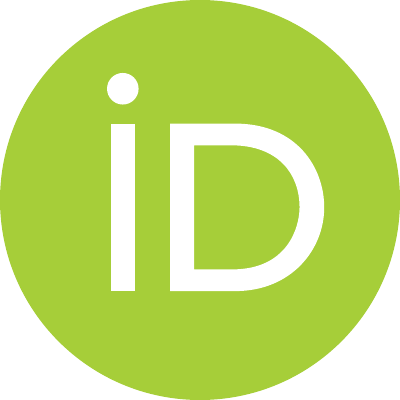}}}}

\begin{document}

%\linenumbers

\title{\boldmath Precision Studies and Searches for CP Asymmetries in the Inclusive Decay $\LcptoLX$}

\author{
\begin{small}
\input{authorlist_2025-09-17.tex}
\end{small}
}

\vspace{4cm}

%\date{\it \small \bf \today}

%%%%%%%%%%%%%%%%
\begin{abstract}
Based on $e^{+}e^{-}$ annihilation data collected with the BESIII detector at center-of-mass energies from $4.600$ to $4.699~\gev$, corresponding to an integrated luminosity of $4.5~\text{fb}^{-1}$, we present the first measurement of the longitudinal polarization of $\Lambda$ hyperon produced in the inclusive decay $\LcptoLX$, where $X$ denotes any allowed final states. The polarizations are determined to be $\mathcal{P}_{\Lambda}=-0.393\pm0.055_{\rm stat}\pm0.020_{\rm syst}$ and $\mathcal{P}_{\bar{\Lambda}}=0.288\pm0.056_{\rm stat}\pm0.017_{\rm syst}$. A search for CP violation is performed using an asymmetry constructed from the $\Lambda$ polarization and the $\Lambda\to p\pi^{-}$ decay asymmetry parameter, and $\acpa=0.15\pm0.12_{\rm stat}\pm0.03_{\rm syst}$ was obtained. Furthermore, an updated measurement of the absolute branching fraction is reported as $\mathcal{B}(\LcptoLX)=(38.07\pm0.38_{\rm stat}\pm0.49_{\rm syst})\%$, representing a fourfold precision improvement over the previous BESIII measurement. A search for CP violation in the decay rate yields $\acpd=(1.5\pm1.0_{\rm stat}\pm1.0_{\rm syst})\%$. No evidence for CP violation is observed in this inclusive charm baryon decay. The subscripts `stat' and `syst' denote statistical and systematic uncertainties, respectively. 
\end{abstract}
%\pacs{14.40.Rt, 13.66.Bc, 13.25.Gv}% PACS, the Physics and Astronomy Classification Scheme.
\maketitle

% --- introduction ---
The standard model (SM) is the most successful and well-tested theory in particle physics, offering a comprehensive framework for describing elementary particles and their fundamental interactions. However, it does not provide a quantitatively sufficient explanation for the observed matter–antimatter asymmetry of the universe.
In 1967, Sakharov proposed that the violation of charge conjugation (C) and charge-parity (CP) symmetry is one of the necessary conditions to explain the baryon asymmetry of the universe~\cite{Sakharov:1991}. 
Over the decades, CP violation (CPV) has been extensively studied in meson systems, first observed in neutral $K$ decays and later established in the $B$ and $D$ sectors~\cite{ref:KsCPV, ref:BCPV_BaBar, ref:BCPV_Belle, LHCb:2019hro}. However, the observable universe is dominated by baryon asymmetry, suggesting that CPV in the baryon sector may play a qualitatively different and potentially critical role.
Recently, the LHCb Collaboration reported the first observation of CPV in the $b$-baryon decay $\Lambda_b^0\to pK^-\pi^+\pi^-$~\cite{LHCb:LbCPV}, with the result consistent with SM predictions. 
By contrast, despite increasing statistics in various experiments, no evidence of CPV has been observed in the charm baryon sector. This is primarily because SM predictions for charm CP asymmetries are suppressed to the order of $\mathcal{O}(10^{-4}\sim10^{-3})$~\cite{Cheng:2012wr}, making such searches experimentally challenging.

As the lightest charm baryon, the $\Lambda_c^+$ serves as a cornerstone for studying weak decays, offering a unique window into the interplay of weak and strong interactions in the baryon sector~\cite{Cheng:2021qpd,Li:2025nzx}. Compared to mesons, the three-quark structure of baryons leads to a richer phenomenology in decay dynamics~\cite{Capstick:2000qj}. Among all studies, inclusive measurements are particularly valuable as they constrain the total decay width and quantify contributions from unmeasured channels. Currently, the sum of all measured exclusive decay modes of $\Lambda_c^+$ containing a $\Lambda$ in the final state accounts for only $(33.59 \pm 1.09)\%$ of the total width~\cite{pdg:2024cfk}. In contrast, the inclusive branching fraction (BF) of $\LcptoLX$ was measured by BESIII to be $(38.2^{+2.8}_{-2.2}\pm0.9)\%$, utilizing a data sample corresponding to $586.9~\rm{pb}^{-1}$~\cite{BESIII:2018Lmdx}. This discrepancy highlights the existence of undiscovered decay modes, but with limited precision. Thus a more accurate inclusive BF measurement is necessary. 

Besides the branching fraction, baryon polarization in weak decays serves as a sensitive probe of parity-violating weak dynamics~\cite{Cheng:2021qpd,Wang:2024wrm}. Previous searches for CPV in charm baryons have focused on exclusive modes, such as $\Lambda_c^+ \to \Lambda K^+$ and $\Lambda_c^+ \to \Sigma^0 K^+$~\cite{LHCb:Lmdpi_alpha,Belle:LmdSgmh}, with results indicating no CPV. While local asymmetries arising from interference in specific exclusive channels may cancel upon integration, a nonzero inclusive asymmetry would indicate a net, integrated difference in decay dynamics between baryons and antibaryons. This motivates inclusive CPV searches as a complementary approach, with the practical advantages of larger signal yields and reduced reliance on modeling individual subchannels~\cite{Wang:2022tcm}. The inclusive decay $\LcptoLX$ has been proposed as a probe of net CPV in the baryon sector~\cite{Wang:2022tcm}, for which SM effects are expected to be $\mathcal{O}(10^{-5}\sim10^{-4})$. 

In this Letter, the inclusive decay $\LcptoLX$ is studied with a double tag (DT) method \cite{ref:DT_method}.
The $\Lcm$ baryon is reconstructed in one of the eleven single tag (ST) modes ($\Modea$, $\Modeb$, $\Modec$, $\Moded$, $\Modee$, $\Modeaa$, $\Modebb$, $\Modedd$, $\Modeaaa$, $\Modeccc$ and $\Modeddd$).
Using  $\ee$ annihilation data, corresponding to an integrated luminosity of 4.5 fb$^{-1}$, collected at center-of-mass energies between 4.600 and 4.699 GeV with the BESIII detector at BEPCII \cite{BESIII:energy2,BESIII:energy3}, we investigate CPV in decay and polarization-dependent CPV in the inclusive decay $\Lambda_c^+ \to \Lambda X$, and report measurements of its BF and the longitudinal polarization of the $\Lambda$ produced in this decay.
Charge-conjugate processes are included unless otherwise stated.

%%%%%%%%%%%%%%%% BESIII and MC simulation %%%%%%%%%%%%%%
Details about BESIII and BEPCII can be found in Refs.~\cite{BESIII:2009fln,ref:14,BESIII:2020nme,Li:2017jpg, Guo:2017sjt, Cao:2020ibk}.
High-statistics inclusive Monte Carlo (MC) simulated samples for $e^+e^-$ annihilation are generated with the {\footnotesize{KKMC}} generator~\cite{ref:Jadach2000ir} incorporating initial-state radiation (ISR) effects and the beam-energy spread. 
The inclusive MC sample, comprising $\Lcp\Lcm$ events and hadronic processes such as $D^{(*)}_{(s)}$ production, ISR return to the lower-mass $\psi$ states, and the continuum processes $\ee\to q \bar{q}$ $(q=u,d,s)$, is generated to determine the ST detection efficiency and estimate the potential background. 
All particle decays are modeled with {\sc evtgen}~\cite{ref:Lange2001uf, ref:Ping2008zz} using BFs either taken from the Particle Data Group (PDG)~\cite{pdg:2024cfk} or estimated with {\sc lundcharm} when available~\cite{Chen:2000tv}.
Final-state particles are propagated through a {\footnotesize{GEANT4}} based detector simulation, which provides an accurate geometric description of the BESIII detector~\cite{GEANT4:2002zbu, Allison:2006ve}.
The Born cross sections are taken into account when generating $\Lcp\Lcm$ pairs in inclusive MC samples~\cite{BESIII:2023rwv}.
Furthermore, signal MC samples are generated to determine the DT detection efficiencies, in which the $\Lcm$ decays into any of the tag modes and the $\Lcp$ decays inclusively into the modes with a $\Lmd$ in the final state. 
These samples incorporate amplitude models for the dominant multibody processes~\cite{BESIII:2022udq} (e.g., $\Lcp \to \Lmd\pi^+\pi^0$) and account for angular distributions in two-body decays~\cite{BESIII:2025zbz,BESIII:xikalpha} (e.g., $\Lcp \to \Lmd\pip$).
To further improve the kinematic agreement between the signal MC sample and data, simulated events are reweighted using a boosted decision tree (BDT)-based algorithm~\cite{hepml}. The BDT is trained using the $\Lambda$ momentum and its recoil invariant mass. The derived weights are then applied in the subsequent analysis.

The selection of ST $\bar{\Lambda}_{c}^{-}$ candidates follows the procedure described in Ref.~\cite{BESIII:nhadron}.
The ST $\bar{\Lambda}_{c}^{-}$ candidates are reconstructed, and the ST yields are extracted from the beam-constrained mass, defined as $M_{\rm BC}\equiv\sqrt{E_{\rm beam}^2/c^4-|\vec{p}^{}|^2/c^2}$, where $E_{\rm beam}$ is the beam energy and $\vec{p}$ is the $\bar{\Lambda}_{c}^-$ momentum in the center-of-mass system of the $e^+e^-$ collision.
\begin{figure*}[!htbp]
	\centering
	\includegraphics[width=0.8\textwidth]{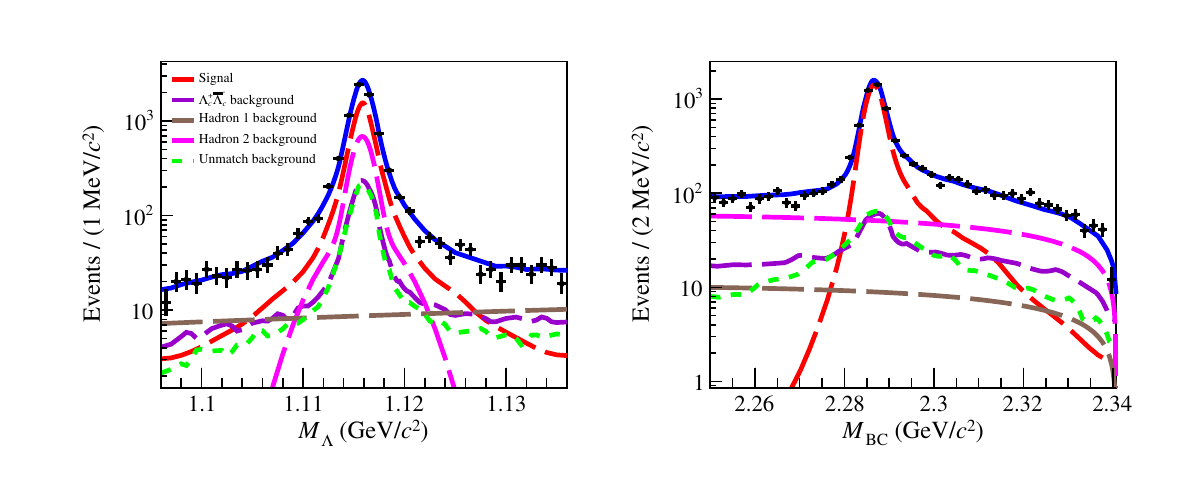}
    \vspace{-7.7mm}
	\caption{Projections of the 2D simultaneous fit on the $M_{\Lmd}$ versus $\mBC$ distributions for the data taken at $\sqrt s=4.682$ GeV, where the vertical axis is shown on a logarithmic scale. Points with error bars are data, and the colored curves show the total fit and its individual components.}
	\label{fig:fit_2D_4680_log}
\end{figure*}
After reconstructing a $\Lcm$ through one of the eleven ST modes, a $\Lmd$ candidate is searched among the remaining tracks via $\Lmdtoppim$, forming the DT candidate. 
The $p\pi^-$ tracks are constrained to originate from a common decay vertex by requiring the $\chi^2$ of the vertex fit to be less than 100. Furthermore, the decay vertex is required to be separated from the interaction point of $\ee$ by a distance of at least twice the fitted vertex resolution.
The squared $\Lambda$ recoil invariant mass distribution for the background-subtracted data sample is shown in Appendix \hyperref[app:lmd_rec]{A}, where a clear peak is observed near zero from the decay $\Lambda_{c}^+\to\Lambda\pi^+$.
To obtain the pure signal shape, a truth-matching procedure is applied to the signal MC samples on both the ST and DT sides.
The matching angle $\theta_{\rm match}$, defined as the angle between the momentum vectors of a reconstructed particle and the corresponding generated particle, must satisfy $\theta_{\rm match}(p)<15^{\circ}$ and $\theta_{\rm match}(\pi^-)<25^{\circ}$ for the $p\pi^-$ pair from the $\Lambda$ in the ST and DT sides, and $\theta_{\rm match}<10^\circ$ is required for other particles in the ST side. Otherwise, they are considered as ``unmatched events''.

To obtain the BF, a two-dimensional (2D) simultaneous unbinned maximum likelihood fit is performed to the distribution of $\mBC$ versus the invariant mass of $p\pim$ ($M_{\Lmd}$) at seven center-of-mass energy points, with all ST modes combined at each energy point.
The signal is modeled by the MC-simulated shape represented by the kernel estimation probability density function (PDF)~\cite{Cranmer:2000du} convolved with a Gaussian function with floating parameters to describe the detector resolution. The unmatched background is described using the shape taken from the signal MC samples. %, also implemented with the one-dimensional kernel estimation PDF. 
The ratio of matched to unmatched event numbers is fixed based on a study of the signal MC samples. 

The background shape from $e^{+}e^{-} \to \Lambda_{c}^{+} \Lcm$ events is determined using inclusive MC samples, which are also used to fix the yield ratio between the signal and this background in the fit.
The background from hadronic processes is classified into two categories: ``Hadron 1,'' which is combinatorial in both the $M_{\Lmd}$ and $\mBC$ distributions, and ``Hadron 2,'' which peaks in $M_{\Lmd}$ distribution but remains combinatorial in $\mBC$ distribution. To model these components, the $M_{\rm BC}$ distributions of both categories are described by ARGUS functions. For the distribution of $M_{\Lmd}$, ``Hadron 1'' is modeled by a polynomial function, whereas ``Hadron 2'' is described by a double Gaussian function. 
The shape parameters are determined by performing the 2D fit to $M_{\Lmd}$ versus $\mBC$ of the hadron MC samples at each energy point, with the ARGUS end points fixed to $E_{\rm beam}$.  
The yields of both components are left as free parameters in the fit.

Figure~\ref{fig:fit_2D_4680_log} shows the fit result of the $M_{\Lmd}$ versus $\mBC$ distribution for data taken at $\sqrt s=4.682$ GeV.
The fit results displayed with linear vertical axes for all energies are shown in Supplemental Material~\cite{supp}.
In the fit, the BF of $\LcptoLX$ is a common parameter at all energy points. The signal yields at each energy point are given as $N^{\rm DT}_{j}=\mathcal{B}\sum_{i} N_{ij}^{\mathrm{ST}} (\epsilon_{ij}^{\mathrm{DT}}/\epsilon_{ij}^{\mathrm{ST}})\mathcal{B}_{\rm int}$,
where $\mathcal{B}$ is the BF of $\LcptoLX$ and $i$ and $j$ represent the ST modes and different center-of-mass energy points, respectively.  
The factor $\mathcal{B}_{\rm int}$ is the BF of $\Lambda\to p\pi^-$, which is $(64.1\pm0.5)\%$ \cite{pdg:2024cfk}.
$N_{ij}^{\mathrm{ST}}$, $\epsilon_{ij}^{\mathrm{ST}}$, and $\epsilon_{ij}^{\mathrm{DT}}$ are the ST yields, ST efficiencies, and DT efficiencies, respectively.
The ST yields and ST efficiencies are cited from Ref.~\cite{BESIII:nhadron} since we use the same reconstruction criteria for ST $\Lcm$ candidates.
The $\epsilon_{ij}^{\mathrm{DT}}$ are estimated from the signal MC samples. 
Furthermore, control samples of $\jpsi\to \bar{p}K^{+}\Lambda$ and $\jpsi\to\Lambda\bar{\Lambda}$ are selected from both data and MC samples to study the $\Lambda$ reconstruction efficiency. Correction factors, parameterized as a function of the $\Lambda$ momentum and $\cos\theta$ (where $\theta$ is the polar angle), are derived from the $\Lambda$ efficiency difference between data and MC samples.
These factors are then applied to correct the DT signal MC samples based on the $\Lambda$ kinematics in the decay $\LcptoLX$. This correction results in a relative difference of $(1.4\pm0.6)\%$ between the corrected and uncorrected efficiencies.
The corrected DT efficiencies for each energy point are summarized in Supplemental Material~\cite{supp}.

In the determination of the BFs, most systematic uncertainties from the ST side cancel out. The remaining systematic uncertainties arise from the $\Lambda$ reconstruction, the intermediate BF of $\Lambda\to p\pi^-$, the 2D fit, the ST $\Lcm$ yield, the BDT reweighting procedure, the MC statistics, and the unmatched background. 
Further details are provided in Appendix \hyperref[app:sys_bf]{B}.
Accounting for these uncertainties, the BF of $\LcptoLX$, averaged over charge-conjugate modes, is determined to be $(38.07\pm0.38_{\rm stat}\pm0.49_{\rm syst})\%$.
The subscripts 'stat' and 'syst' denote statistical and systematic uncertainties, respectively. 
Applying the same procedure to the individual charge-conjugate modes, we determine the BFs for the $\LcptoLX$ and $\LcmtoLX$ decays to be $(38.35\pm0.53_{\rm stat}\pm0.69_{\rm syst})\%$ and $(37.19\pm0.53_{\rm stat}\pm0.60_{\rm syst})\%$, respectively.
The ST yields, ST efficiencies, and DT efficiencies for the separate decays are summarized in Supplemental Material~\cite{supp}.

\begin{figure}[htbp]
	\centering
	\includegraphics[width=0.42\textwidth]{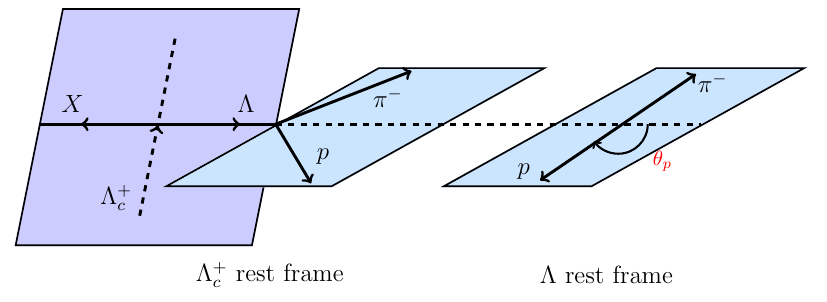}
    \vspace{-4.5mm}
	\caption{Angular distribution of the inclusive decay $\LcptoLX$ and its subsequent decay $\Lmd\to p\pim$.}
	\label{fig:ang_dis}
\end{figure}

% --- polarization parameters analysis ---
The longitudinal polarization of the $\Lambda$ hyperon is extracted from the proton angular distribution in the decay $\Lcp\to\Lambda X$ with $\Lambda\to p\pi^-$. The decay topology and the definition of the reference frames are illustrated in Fig.~\ref{fig:ang_dis}.
As described in Ref.~\cite{Wang:2022tcm}, the differential decay rate for this decay can be expressed in terms of a single helicity angle:
\begin{equation}
\label{eq:ang_fuc}
\frac{d\Gamma}{d\cos\theta_{p}}\propto 1+\mathcal{P}_{\Lambda}\alpha_{-}\cos\theta_{p},
\end{equation}
where $\alpha_{-}$ is the decay asymmetry parameter for $\Lambda$ in $\Lambda\to p\pi^-$.
The helicity axis is defined along the $\Lambda$ flight direction in the $\Lambda_c^+$ rest frame. The helicity angle, $\theta_{p}$, is defined as the angle between the proton momentum and this axis, measured in the $\Lambda$ rest frame.
The parameter $\mathcal{P}_{\Lambda}$ denotes the longitudinal polarization of the $\Lambda$ hyperon
along the chosen helicity axis, averaged over all contributing $\Lambda_c^+\to\Lambda X$ decay modes.
For the charge-conjugate decay $\LcmtoLX$, the $\alpha_{-}$ and $\mathcal{P}_{\Lambda}$ are replaced by $\alpha_{+}$ and $\mathcal{P}_{\bar{\Lambda}}$, respectively.

\begin{figure}[hbtp]
	%\centering
	{
		\includegraphics[width=0.4\textwidth]{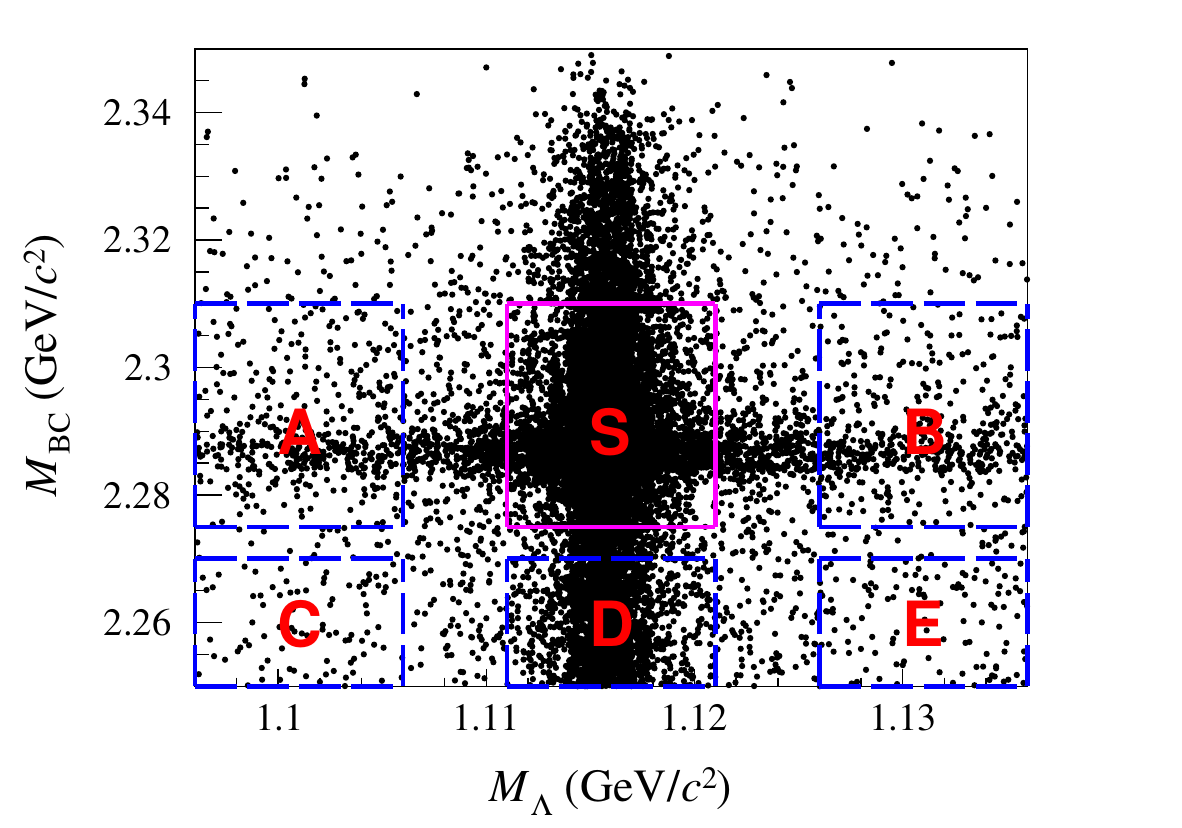}
	}
    \vspace{-3.8mm}
	\caption{Distribution of $\mBC$ versus $M_{\Lmd}$ in data taken at $\sqrt{s}=4.682\gev$. The $\Lambda$ signal region is defined as (1.111, 1.121) $\gevcc$, with sidebands taken as (1.096, 1.106) $\cup$ (1.126, 1.136) $\gevcc$, while the $\mBC$ signal and sideband regions are chosen as (2.275, 2.31) $\gevcc$ and (2.25, 2.27) $\gevcc$, respectively.
	}
	\label{fig:data2D}
\end{figure}

% --- fitting for polarization parameter ---
A simultaneous unbinned maximum likelihood fit for the seven data samples is performed to determine the common parameters.
The total effective log-likelihood function is given by $\ln\mathcal{L}^{\text{sig}}_{\text{total}}=\sum\limits^{\text{energy}}\ln\mathcal{L}^{\text{sig}}_{\text{energy}}$.
At each energy point, the background contribution is statistically subtracted at the log-likelihood level~\cite{BESIII:2022udq} according to 
\begin{equation}
    \label{eq:loglikelihood}
    \begin{aligned}
    \ln\mathcal{L}_{\text{energy}}^{\text{sig}}
    &=\kappa\Biggl[\sum_{i\in \rm S}\ln f_{s}(\theta_{p}^{i};\eta)
    -\frac{1}{2}\sum_{i\in \rm A\cup B}\ln f_{s}(\theta_{p}^{i};\eta) \\
    &-r\sum_{i\in \rm D}\!\ln f_{s}(\theta_{p}^{i};\eta)
    +\frac{r}{2}\sum_{i\in \rm C\cup E}\!\ln f_{s}(\theta_{p}^{i};\eta)\Biggr].
    \end{aligned}
\end{equation}
%\begin{equation}
%    \label{eq:loglikelihood}
%    \ln\mathcal{L}_{\text{energy}}^{\text{sig}}=\kappa\cdot\left[\sum_{i\in \rm S}\ln f_{s}(\theta_{p}^{i};\eta)-r\cdot\sum_{i\in \rm sideband}\ln f_{s}(\theta_{p}^{i};\eta)\right].
%\end{equation}
The factor $\kappa = \frac{N_{\rm S}-(N_{\rm A}+N_{\rm B})/2-r[N_{\rm D}-(N_{\rm C}+N_{\rm E})/2]}{N_{\rm S}+(N_{\rm A}+N_{\rm B})/4+r^2[N_{\rm D}+(N_{\rm C}+N_{\rm E})/4]}$ accounts for the event weights in the effective
log-likelihood and corrects the parameter uncertainties following Refs.~\cite{BESIII:2022udq,Langenbruch:2019nwe}, where $N_j$ is the data yield in region $j$ ($j\in\{S,A,B,C,D,E\}$). %, where $N_{\rm data}$ is the number of events in the signal region, $N_{\rm bkg}$ is the estimated background yield from the sideband region, defined as $N_{\rm bkg}=(N^A_{\rm data}+N^B_{\rm data})/2+N^D_{\rm data}-(N^C_{\rm data}+N^E_{\rm data})/2$, and $V_{\rm bkg}$ is the statistical variance of the $N_{\rm bkg}$, given by $V_{\rm bkg}=(N^A_{\rm data}+N^B_{\rm data})/4+N^D_{\rm data}+(N^C_{\rm data}+N^E_{\rm data})/4$.
As shown in Fig.~\ref{fig:data2D}, the box labeled $S$ stands for the signal region, and boxes $A$, $B$, $C$, $D$ and $E$ represent the 2D sideband regions.
The background normalization ratio $r$ between the $\mBC$ sideband and signal regions is determined from the 2D simultaneous fit.
The $\cos\theta_p$ distribution of the data sideband events is corrected
for the difference between the background distributions in the signal
and sideband regions, as determined from background MC samples including
$\Lcp\Lcm$ pairs, hadronic processes, and unmatched components.
%A correction is applied to the $\cos\theta_{p}$ distribution of the data sideband events, accounting for the observed difference between the signal and the sideband regions in the background MC samples, which include contributions from $\Lcp\Lcm$ pairs, hadronic processes, and unmatched components.
In the $\mBC$ distribution, only the lower sidebands are used, since the high side ISR tail contains signal events that would contaminate the background estimate.
%%%%%%%%%%
%The likelihood terms are constructed as  $\mathcal{L}^{\text{data(bkg)}}_{\text{energy}}=\prod\limits_{i=1}\limits^{N_{\text{data(bkg)}}}f_{s}(\theta_{p}^{i};\eta)$.
%Here, $i$ labels the event. 
%\begin{equation}
%    \label{likelihood}
%    L^{\text{data\/bkg}}_{\text{energy}}=\prod_{i=1}^{N_{\text{data}}}f_{s}(\theta_{p}^{i};\eta).
%\end{equation}
The PDF is formulated as $f_{s}(\theta_{p}^{i};\eta)=\frac{\epsilon(\theta_{p}^{i})|M(\theta_{p}^{i};\eta)|^{2}}{\int\epsilon(\theta_{p}^{i})|M(\theta_{p}^{i};\eta)|^{2}\text{d}\theta_{p}^{i}}$,
where $M(\theta_{p}^{i};\eta)$ is the angular distribution function described in Eq.~\eqref{eq:ang_fuc}, $\epsilon(\theta_{p}^{i})$ is the detection efficiency parameterized in terms of the kinematic variables $\theta_{p}^{i}$, and $\eta$ denotes the product $\mathcal{P}_{\Lambda}\alpha_{-}$ or $\mathcal{P}_{\bar{\Lambda}}\alpha_{+}$ to be determined.
The integration of the normalization factor in the fitting procedure is calculated using a large BDT reweighted signal MC sample, with $\int\epsilon(\theta_{p}^{i})|M(\theta_{p}^{i};\eta)|^{2}\text{d}\theta_{p}^{i}\propto\frac{1}{N_{\text{gen}}}\sum_{k_{\text{MC}}}^{N_{\text{MC}}}|M(\theta_{p}^{k_{\text{MC}}};\eta)|^{2}$, 
where $N_{\text{gen}}$ is the total number of the simulated signal MC events, $N_{\text{MC}}$ is the number of the reweighted signal MC events surviving all selection criteria and $k_{\text{MC}}$ is the event index.
\begin{figure*}[hpbt]
	\centering
	\includegraphics[width=0.85\textwidth]{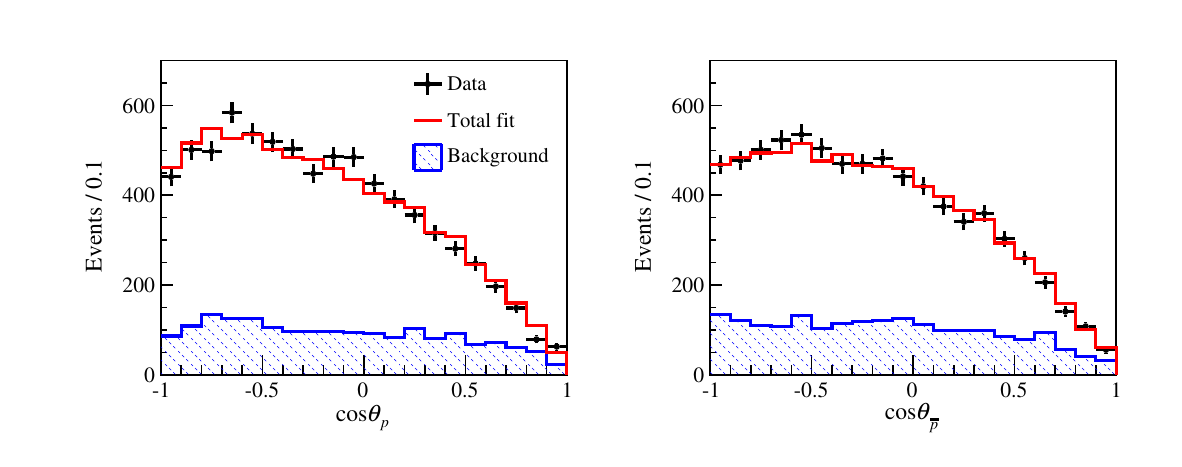}
	\vspace{-7.7mm}
	\caption{Angular distributions for (left) $\LcptoLX$ and (right) $\LcmtoLX$, obtained by summing over the data samples of all the energy points. Points with error bars are data, the red lines are the fitting models, and the blue shaded areas are the background events of the 2D sideband region in data.
    }
	\label{fig:anglefit}
\end{figure*}
The negative log-likelihood with background subtraction over the seven data samples is minimized using the MINUIT package~\cite{James:1975dr}.
Figure~\ref{fig:anglefit} shows the fit results for the decays $\LcptoLX$ and $\LcmtoLX$. 
For illustration, the corresponding background-subtracted and
efficiency-corrected angular distributions are presented in Supplemental Material~\cite{supp}.

Systematic uncertainties from imperfect simulation and the measurement procedure can affect the determination of the products $\mathcal{P}_{\Lambda}\alpha_{-}$ and $\mathcal{P}_{\bar{\Lambda}}\alpha_{+}$. The relevant sources are the $\Lambda$ reconstruction, the resolution differences between data and MC simulation, the background subtraction, the unmatched-background model, and the BDT reweighting procedure. These sources also apply to the $\Lambda$ longitudinal polarizations $\mathcal{P}_{\Lambda}$ and $\mathcal{P}_{\bar{\Lambda}}$. The additional systematic uncertainty for the $\mathcal{P}_{\Lambda}$ and $\mathcal{P}_{\bar{\Lambda}}$ arises from the input parameters $\alpha_{-,+}$, which do not affect the products extracted directly from data. 
The BESIII inputs to the PDG averages of $\alpha_{-,+}$ are based on the $J/\psi$ data and do not overlap with the data used in the present analysis. 
This uncertainty, estimated by Gaussian sampling according to the uncertainties of $\alpha_{-,+}$, is $1.0\%$ for $\mathcal{P}_{\Lambda}$ and $0.7\%$ for $\mathcal{P}_{\bar{\Lambda}}$. 
Further details are given in Appendix~\hyperref[app:sys_pol]{C}.
The measured products are $\mathcal{P}_{\Lambda}\alpha_{-}=-0.293\pm0.041_{\rm stat}\pm0.015_{\rm syst}$ and $\mathcal{P}_{\bar{\Lambda}}\alpha_{+}=-0.218\pm0.042_{\rm stat}\pm0.013_{\rm syst}$.
Using the PDG values $\alpha_{-}=0.746\pm0.008$ and $\alpha_{+}=-0.758\pm0.005$~\cite{pdg:2024cfk}, 
the $\Lambda$ longitudinal polarizations are determined to be $\mathcal{P}_{\Lambda}=-0.393\pm0.055_{\rm stat}\pm0.020_{\rm syst}$ and $\mathcal{P}_{\bar{\Lambda}}=0.288\pm0.056_{\rm stat}\pm0.017_{\rm syst}$, respectively. 

%%%%%%%%%%%%%%%%%%% direct CPV and alpha-induced CPV %%%%%%%%%%%%%%
The CP asymmetry between $\LcptoLX$ and $\LcmtoLX$, defined as $\acpd\equiv[\mathcal{B}(\LcptoLX)-\mathcal{B}(\LcmtoLX)]/[\mathcal{B}(\LcptoLX)+\mathcal{B}(\LcmtoLX)]$, is determined to be $\acpd=(1.5\pm1.0_{\rm stat}\pm1.0_{\rm syst})\%$.
The CP asymmetry constructed from the products $\mathcal{P}_{\Lambda}\alpha_{-}$ and $\mathcal{P}_{\bar{\Lambda}}\alpha_{+}$, $\acpa\equiv(\mathcal{P}_{\Lambda}\alpha_{-}-\mathcal{P}_{\bar{\Lambda}}\alpha_{+})/(\mathcal{P}_{\Lambda}\alpha_{-}+\mathcal{P}_{\bar{\Lambda}}\alpha_{+})$, provides access to polarization-dependent CPV in the inclusive decay $\LcptoLX$ and is measured to be $\acpa=0.15\pm0.12_{\rm stat}\pm0.03_{\rm syst}$.
This is the first search for polarization-dependent CPV in the inclusive decay $\LcptoLX$. 
The corresponding systematic uncertainty evaluations for $\acpd$ and $\acpa$ are summarized in Appendixes~\hyperref[app:sys_bf]{B} and \hyperref[app:sys_pol]{C}, respectively.

In summary, using data corresponding to an integrated luminosity of 4.5 $\rm{fb}^{-1}$, collected by BESIII at center-of-mass energies between 4.600 and 4.699 GeV, we report the first measurement of the longitudinal polarizations of $\Lambda$ hyperon produced in inclusive decay $\LcptoLX$, which are determined to be $\mathcal{P}_{\Lambda}=-0.393\pm0.055_{\rm stat}\pm0.020_{\rm syst}$ and $\mathcal{P}_{\bar{\Lambda}}=0.288\pm0.056_{\rm stat}\pm0.017_{\rm syst}$.
These results provide detailed information that improves our understanding of the dynamics of charm baryon decay.
The absolute BF of the inclusive decay $\LcptoLX$ is determined to be $\mathcal{B}(\LcptoLX)=(38.07\pm0.38_{\rm stat}\pm0.49_{\rm syst})\%$, with precision improved by a factor of four over the previous BESIII measurement~\cite{BESIII:2018Lmdx}. 
This more precise determination constrains the sum of known exclusive $\Lambda_c^+$ decay channels, thereby reducing the allowed phase space for yet-undiscovered modes and providing a clear target for future searches.
By comparison with the measured exclusive processes containing a $\Lmd$ in the final state listed in the PDG~\cite{pdg:2024cfk}, it is found that $(4.48\pm1.25)$\% of the exclusive decays remain unmeasured.
Moreover, we measure the CP asymmetry in decay to be $\acpd=(1.5\pm1.0_{\rm stat}\pm1.0_{\rm syst})\%$. 
The present analysis uses the data that includes the one used in the previous BESIII measurement and additional ST modes, and therefore the results of the BF and $\acpd$ supersede the previous BESIII results.
We also perform the first search for the polarization-dependent CPV in this inclusive decay, obtaining $\acpa = 0.15 \pm 0.12_{\rm stat} \pm 0.03_{\rm syst}$. 
By searching for the CPV in inclusive mode, we provide a complementary constraint on a net CPV in $\LcptoLX$ decay that survives after integration over all contributing subchannels, which is relevant for understanding the mechanisms underlying the baryon asymmetry of the universe.
These results offer new insights to improve our understanding of CPV in charm baryons and the properties of inclusive decays.

%%%%%%%%%%%%%%%%%%%%%%%%%%%%%%%%%%%%%%%%%%%%%%%%%%%%%%%%%%%%%%%%
%%%%%     Acknowledgments                          %%%%%%%%%%%%%
%%%%%%%%%%%%%%%%%%%%%%%%%%%%%%%%%%%%%%%%%%%%%%%%%%%%%%%%%%%%%%%%
\acknowledgments
\input{acknowledgement_2025-09-17.tex}

\input{bibitem}
%%%%%%%%%%%%%%%%%%%%%%%%%%%%%%%%%%%%%%%%%%%%%%%%%%%%%%%%%%%%%%%%
%%%%%     End Matter                          %%%%%%%%%%%%%
%%%%%%%%%%%%%%%%%%%%%%%%%%%%%%%%%%%%%%%%%%%%%%%%%%%%%%%%%%%%%%%%
\onecolumngrid
\begin{center}
\textbf{\boldmath\large End Matter}
\end{center}
\twocolumngrid
\appendix*
\setcounter{equation}{0}
\renewcommand{\theequation}{A\arabic{equation}}

\begin{figure}[!h]
	\centering
	\includegraphics[width=0.4\textwidth]{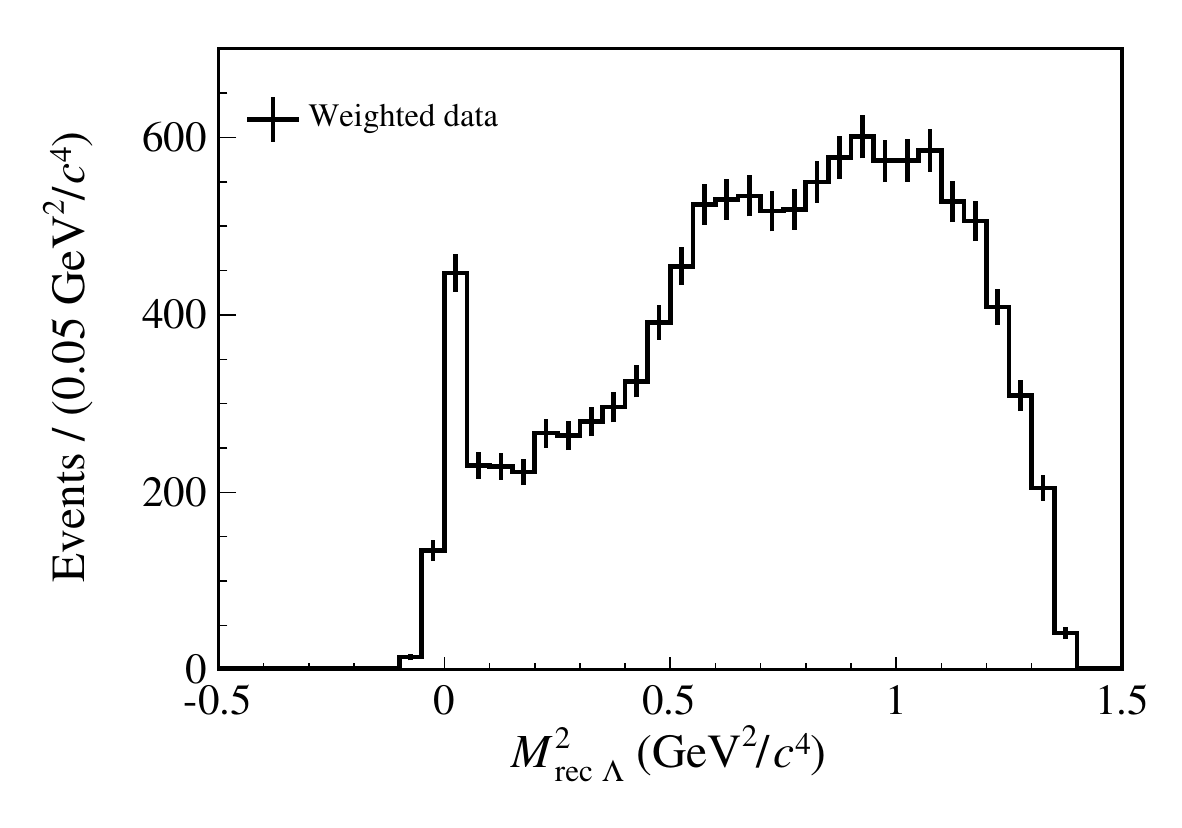}
	\vspace{-6.5mm}
	\caption{Squared $\Lambda$ recoil invariant mass distribution of the weighted data sample.}
	\label{fig:lmd_rec_mass}
\end{figure}

\phantomsection
\label{app:lmd_rec}
\textit{Appendix A: Plot for the squared $\it \Lambda$ recoil invariant mass distribution}---Figure~\ref{fig:lmd_rec_mass} shows the squared $\Lambda$ recoil invariant mass distribution of the weighted data sample.
The clear peak above zero corresponds to the $\Lambda_{c}^+\to\Lambda\pi^+$ decay.

\phantomsection
\label{app:sys_bf}
\textit{Appendix B: Systematic uncertainties for the BF and $\acpd$}---The systematic uncertainty associated with the $\Lambda$ reconstruction is derived from the uncertainty on the efficiency correction factor, which is estimated using the control samples $J/\psi\to \bar{p}K^+\Lambda$ and $J/\psi\to\Lambda\bar{\Lambda}$. Additionally, the uncertainty related to the intermediate BF of $\Lambda\to p\pi^-$ is taken from the PDG.
For the 2D fit, systematic uncertainty is evaluated by varying the signal shape, including a convolution of the MC-simulated shape with a double-Gaussian function, and by varying the background shape and the ratio of the signal and $\Lcp\Lcm$ background yields.
Similarly, the uncertainty on the total ST $\Lcm$ yield is assessed from variations of the signal and background shapes as well as the fit range in the $M_{\rm BC}$ fit. 
To evaluate the uncertainty from the BDT reweighting procedure, the BDT hyperparameters are varied, and the corresponding uncertainty is estimated from the resulting efficiency differences between the nominal and modified samples. Limited MC statistics contribute an additional uncertainty, quantified through sample size fluctuations. For the unmatched background, the corresponding uncertainty is assigned based on the deviation observed in the BF when relaxing the matching angle requirement.

The uncertainties of $\acpd$ are obtained by propagating the BF uncertainties for the decays $\Lambda_{c}^{+}\to\Lambda X$ and $\bar{\Lambda}_{c}^{-}\to\bar{\Lambda} X$. 
In this propagation, the uncertainty associated with the intermediate BF of $\Lambda\to p\pi^-$ is treated as fully correlated, since possible CPV in this decay is neglected and the same PDG value is used for both charge-conjugate modes. The contribution from the BDT reweighting procedure is also taken to be fully correlated, because the same set of BDT hyperparameters is used in evaluating this effect for the two charge-conjugate modes.
By contrast, the uncertainty associated with the $\Lambda$ reconstruction is taken from the uncertainties of the efficiency correction factors, which are determined by the statistical precision of the respective control samples, and is therefore treated as uncorrelated. 
The remaining sources are evaluated independently using separate samples for the decays $\Lambda_{c}^{+}\to\Lambda X$ and $\bar{\Lambda}_{c}^{-}\to\bar{\Lambda} X$ and are thus also treated as uncorrelated in the present analysis.

The total systematic uncertainty is taken as the quadratic sum of the individual contributions. The corresponding systematic uncertainties for the CP-averaged BF, the charge-conjugate BFs, and $\acpd$ are summarized in Table~\ref{tab:sys_err_bf_acp}.
\begin{table}[!htbp]
    \centering
    \caption{Systematic uncertainties (in percent) for the CP averaged and
charge-conjugate BFs and $\acpd$, given as relative
uncertainties for the BFs and absolute uncertainties for
$\acpd$.}
    \setlength{\tabcolsep}{4pt}
    \renewcommand{\arraystretch}{1.18}
    \begin{tabular}{l
                    S[table-format=1.1]
                    S[table-format=1.1]
                    S[table-format=1.1]
                    S[table-format=1.1]}
        \hline
        Source
        & {CP average}
        & {$\Lambda X$}
        & {$\bar{\Lambda}X$}
        & {$\acpd$} \\
        \hline
        $\Lambda$ reconstruction & 0.6 & 0.7 & 0.7 & 0.5 \\
        Intermediate BF          & 0.8 & 0.8 & 0.8 & 0.0 \\
        2D fit                   & 0.4 & 0.9 & 0.6 & 0.4 \\
        ST $\Lcm$ yield          & 0.4 & 0.5 & 0.6 & 0.4 \\
        BDT reweighting          & 0.1 & 0.1 & 0.1 & 0.0 \\
        MC statistics            & 0.2 & 0.3 & 0.3 & 0.2 \\
        Unmatched background     & 0.6 & 1.0 & 0.8 & 0.6 \\
        \hline
        Total                    & 1.3 & 1.8 & 1.6 & 1.0 \\
        \hline
    \end{tabular}
    \label{tab:sys_err_bf_acp}
\end{table}

\phantomsection
\label{app:sys_pol}
\textit{Appendix C: Systematic uncertainties for the products $\mathcal{P}_{\Lambda}\alpha_{-}$ and $\mathcal{P}_{\bar{\Lambda}}\alpha_{+}$ and for $\acpa$}---The systematic uncertainty arising from the $\Lmd$ reconstruction is evaluated using the control samples $J/\psi\to \bar{p}K^+\Lambda$ and $J/\psi\to\Lambda\bar{\Lambda}$. 
Correction factors extracted from these control samples are applied to the signal MC samples, modifying the
$\theta_p$-dependent efficiency used in the likelihood normalization and
consequently leading to changes in the fit results.
The systematic uncertainty for the differences in the
$M_{\rm BC}$ and $M_{\Lambda}$ resolutions between data and signal MC
is evaluated by applying Gaussian smearing to the $M_{\rm BC}$ and $M_{\Lambda}$ variables in the signal MC samples, using
smearing parameters determined from fits to the corresponding
distributions, and repeating the MC integration in the angular distribution
fit.
The systematic uncertainty associated with the BDT reweighting procedure is obtained by changing the BDT hyperparameters.
For the background subtraction, we take into account the background shape and size.
The systematic uncertainty associated with the background shape is evaluated by replacing the data sideband sample with the background MC sample from the signal region. 
To assess the impact of the background size, the fits are repeated using alternative background yields derived from Gaussian sampling of the 2D fit parameters. By sampling the fit parameters under the assumption of a Gaussian distribution, we estimate the systematic uncertainty associated with the background size. The uncertainty is evaluated by combining the fitted Gaussian width and shift of the fitted Gaussian mean from the nominal value.
In addition, to account for potential mismodeling of the unmatched background component, an alternative signal MC sample is generated using polarizations shifted by one statistical standard deviation from their nominal values, and the resulting variation is assigned as the corresponding systematic uncertainty.

The uncertainties of $\acpa$ are propagated from those of the products $\mathcal{P}_{\Lambda}\alpha_{-}$ and $\mathcal{P}_{\bar{\Lambda}}\alpha_{+}$.
The contribution associated with the BDT reweighting procedure is treated as fully correlated, following the same treatment as for $\acpd$.
The $\Lambda$ reconstruction uncertainty is treated as partially correlated between the two charge-conjugate modes, since the determination of the corresponding efficiency correction factors contain both common and channel-specific contributions. 
The common contributions arise from three correlated terms: the reconstruction of charged tracks, the particle identification and the vertex fit; the channel-specific contributions come from two uncorrelated terms: the finite size of control samples and the extraction of yields. Therefore, a correlation coefficient of $60\%$ is adopted when evaluating the uncertainty on $\acpa$. As a robustness check, the correlation coefficient is varied to 50\% and 70\%, and the final result remains unchanged at the quoted precision. For other contributions, the corresponding uncertainties are determined separately for the decays $\Lambda_{c}^{+}\to\Lambda X$ and $\bar{\Lambda}_{c}^{-}\to\bar{\Lambda} X$, and are therefore taken to be uncorrelated.

All systematic uncertainties are added in quadrature to calculate the total systematic uncertainties.
The corresponding uncertainties for the products $\mathcal{P}_{\Lambda}\alpha_{-}$ and $\mathcal{P}_{\bar{\Lambda}}\alpha_{+}$ and for $\acpa$ are summarized in Table~\ref{tab:sysplor}.
\begin{table}[!hb]
	\centering
	\caption{Relative systematic uncertainties (in percent) for the products $\mathcal{P}_{\Lambda}\alpha_{-}$ and for $\mathcal{P}_{\bar{\Lambda}}\alpha_{+}$ and $\acpa$. }
	\setlength{\tabcolsep}{6pt}
	\renewcommand{\arraystretch}{1.18}
	\begin{tabular}{l
			S[table-format=1.1]
			S[table-format=1.1]
			S[table-format=2.1]}
		\hline
		Source & {$\mathcal{P}_{\Lambda}\alpha_{-}$} & {$\mathcal{P}_{\bar{\Lambda}}\alpha_{+}$} & {$\acpa$} \\
		\hline
		$\Lambda$ reconstruction & 4.6 & 5.2 & 14.7 \\
		Resolution               & 0.3 & 0.3 & 1.4  \\
		BDT reweighting          & 0.5 & 0.3 & 0.5  \\
		Background size          & 1.5 & 1.4 & 6.9  \\
		Background shape         & 0.3 & 2.4 & 8.1  \\
		Unmatched background     & 1.0 & 1.0 & 4.9  \\
		%Input parameter          & 1.0 & 0.7 & \multicolumn{1}{r}{$\cdots$} \\
		\hline
		Total                    & 5.0 & 6.0 & 18.9 \\
		\hline
	\end{tabular}
	\label{tab:sysplor}
\end{table}

%Supplemental Material

\onecolumngrid
\clearpage
%\appendix

\begin{center}
\textbf{\boldmath\large Supplemental Material for ``Precision Studies and Searches for CP Asymmetries in the Inclusive Decay $\LcptoLX$"}
\end{center}

\begin{center}
\begin{small}
\input{authorlist_2025-09-17.tex}
\end{small}
\end{center}

\newpage

\input{Supplementary_Lc2LmdX_arXiv.tex}

\end{document}

%% file: authorlist_2025-09-17.tex
M.~Ablikim$^{1}$\BESIIIorcid{0000-0002-3935-619X},
M.~N.~Achasov$^{4,c}$\BESIIIorcid{0000-0002-9400-8622},
P.~Adlarson$^{81}$\BESIIIorcid{0000-0001-6280-3851},
X.~C.~Ai$^{87}$\BESIIIorcid{0000-0003-3856-2415},
C.~S.~Akondi$^{31A,31B}$\BESIIIorcid{0000-0001-6303-5217},
R.~Aliberti$^{39}$\BESIIIorcid{0000-0003-3500-4012},
A.~Amoroso$^{80A,80C}$\BESIIIorcid{0000-0002-3095-8610},
Q.~An$^{77,64,\dagger}$,
Y.~H.~An$^{87}$\BESIIIorcid{0009-0008-3419-0849},
Y.~Bai$^{62}$\BESIIIorcid{0000-0001-6593-5665}
% ...
\\
\vspace{0.2cm}
(BESIII Collaboration)\\
\vspace{0.2cm} {\it
$^{1}$ Institute of High Energy Physics, Beijing 100049, People's Republic of China\\
% ...
$^{4}$ Budker Institute of Nuclear Physics SB RAS (BINP), Novosibirsk 630090, Russia\\
% ...
$^{31}$ INFN Sezione di Ferrara, (A)INFN Sezione di Ferrara, I-44122, Ferrara, Italy; (B)University of Ferrara, I-44122, Ferrara, Italy\\
% ...
$^{39}$ Johannes Gutenberg University of Mainz, Johann-Joachim-Becher-Weg 45, D-55099 Mainz, Germany\\
% ...
$^{62}$ Southeast University, Nanjing 211100, People's Republic of China\\
% ...
$^{64}$ State Key Laboratory of Particle Detection and Electronics, Beijing 100049, Hefei 230026, People's Republic of China\\
% ...
$^{77}$ University of Science and Technology of China, Hefei 230026, People's Republic of China\\
% ...
$^{80}$ University of Turin and INFN, (A)University of Turin, I-10125, Turin, Italy; (B)University of Eastern Piedmont, I-15121, Alessandria, Italy; (C)INFN, I-10125, Turin, Italy\\
$^{81}$ Uppsala University, Box 516, SE-75120 Uppsala, Sweden\\
% ...
$^{87}$ Zhengzhou University, Zhengzhou 450001, People's Republic of China\\

\vspace{0.2cm}
$^{\dagger}$ Deceased\\
% ...
$^{c}$ Also at the Novosibirsk State University, Novosibirsk, 630090, Russia\\
% ...
}

%% file: acknowledgement_2025-09-17.tex
%% Saved at => 2025-09-17
%\textbf{Acknowledgement}

The BESIII Collaboration thanks the staff of BEPCII (https://cstr.cn/31109.02.BEPC) and the IHEP computing center for their strong support. This work is supported in part by National Key R\&D Program of China under Contracts Nos. 2023YFA1609400, 2023YFA1606000, 2023YFA1606704; National Natural Science Foundation of China (NSFC) under Contracts Nos. 12105127, 12422504, 11635010, 11935015, 11935016, 11935018, 12025502, 12035009, 12035013, 12061131003, 12192260, 12192261, 12192262, 12192263, 12192264, 12192265, 12221005, 12225509, 12235017, 12342502, 12361141819; the Chinese Academy of Sciences (CAS) Large-Scale Scientific Facility Program; the Strategic Priority Research Program of Chinese Academy of Sciences under Contract No. XDA0480600; CAS under Contract No. YSBR-101; 100 Talents Program of CAS; Fundamental Research Funds for the Central Universities, Lanzhou University under Contracts Nos. lzujbky-2025-ytB01, lzujbky-2023-stlt01; The Institute of Nuclear and Particle Physics (INPAC) and Shanghai Key Laboratory for Particle Physics and Cosmology; ERC under Contract No. 758462; German Research Foundation DFG under Contract No. FOR5327; Istituto Nazionale di Fisica Nucleare, Italy; Knut and Alice Wallenberg Foundation under Contracts Nos. 2021.0174, 2021.0299, 2023.0315; Ministry of Development of Turkey under Contract No. DPT2006K-120470; National Research Foundation of Korea under Contract No. NRF-2022R1A2C1092335; National Science and Technology fund of Mongolia; Polish National Science Centre under Contract No. 2024/53/B/ST2/00975; STFC (United Kingdom); Swedish Research Council under Contract No. 2019.04595; U. S. Department of Energy under Contract No. DE-FG02-05ER41374

%\textbf{Other Fund Information}
%
%To be inserted with an additional sentence into papers that are relevant to the topic of special funding for specific topics. Authors can suggest which to Li Weiguo and/or the physics coordinator.
%        Example added sentence: This paper is also supported by the NSFC under Contract Nos. 10805053, 10979059, ....National Natural Science Foundation of China (NSFC), 10805053, PWANational Natural Science Foundation of China (NSFC), 10979059, Lund弦碎裂强子化模型及其通用强子产生器研究National Natural Science Foundation of China (NSFC), 10775075, National Natural Science Foundation of China (NSFC), 10979012, baryonsNational Natural Science Foundation of China (NSFC), 10979038, charmoniumNational Natural Science Foundation of China (NSFC), 10905034, psi(2S)->B BbarNational Natural Science Foundation of China (NSFC), 10975093, D 介子National Natural Science Foundation of China (NSFC), 10979033, psi(2S)->VPNational Natural Science Foundation of China (NSFC), 10979058, hcNational Natural Science Foundation of China (NSFC), 10975143, charmonium rare decays
%%% ends here %%

%% file: Supplementary_Lc2LmdX_arXiv.tex
\section{Plots for the various energies}

\begin{figure*}[!htb]
	\centering
	\includegraphics[width=0.8\textwidth]{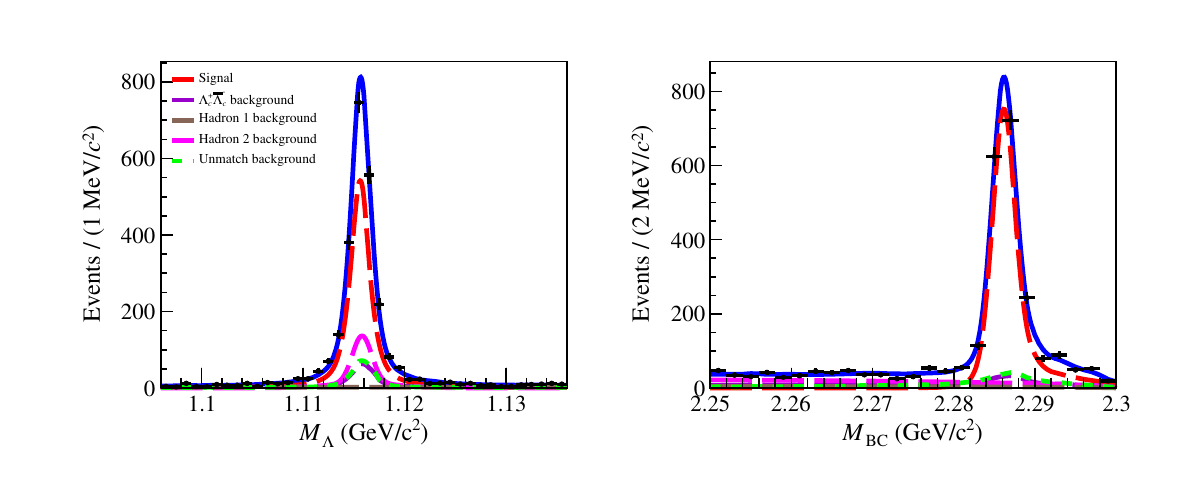}
	\vspace{-7.7mm}
	\caption{Projections of the 2D simultaneous fit on the $M_{\Lmd}$ versus $\mBC$ distributions for the data taken at $\sqrt s=4.600$ GeV. Points with error bars are data, and the colored curves show the total fit and its individual components.%Points with error bars are data, the blue solid lines are the sum of fitting functions, the red dashed lines are the $\LcptoLX$ signal, the purple dashed lines are the $\Lcp\Lcm$ background, the brown dashed lines are the ``hadron 1'' background, the magenta dashed lines are the ``hadron 2'' background and the green dashed lines are the unmatched background.
    }
	\label{fig:fit_2D_4600}
\end{figure*}

\begin{figure*}[!htb]
	\centering
	\includegraphics[width=0.8\textwidth]{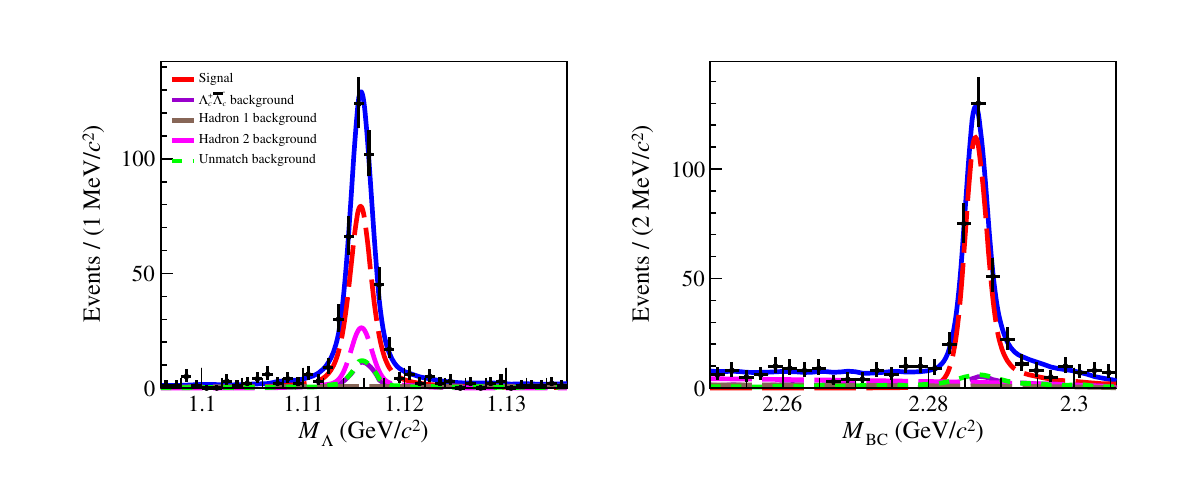}
	\vspace{-7.7mm}
	\caption{Projections of the 2D simultaneous fit on the $M_{\Lmd}$ versus $\mBC$ distributions for the data taken at $\sqrt s=4.612$ GeV. Points with error bars are data, and the colored curves show the total fit and its individual components.%Points with error bars are data, the blue solid lines are the sum of fitting functions, the red dashed lines are the $\LcptoLX$ signal, the purple dashed lines are the $\Lcp\Lcm$ background, the brown dashed lines are the ``hadron 1'' background, the magenta dashed lines are the ``hadron 2'' background and the green dashed lines are the unmatched background.
    }
	\label{fig:fit_2D_4612}
\end{figure*}

\begin{figure*}[!htb]
	\centering
	\includegraphics[width=0.8\textwidth]{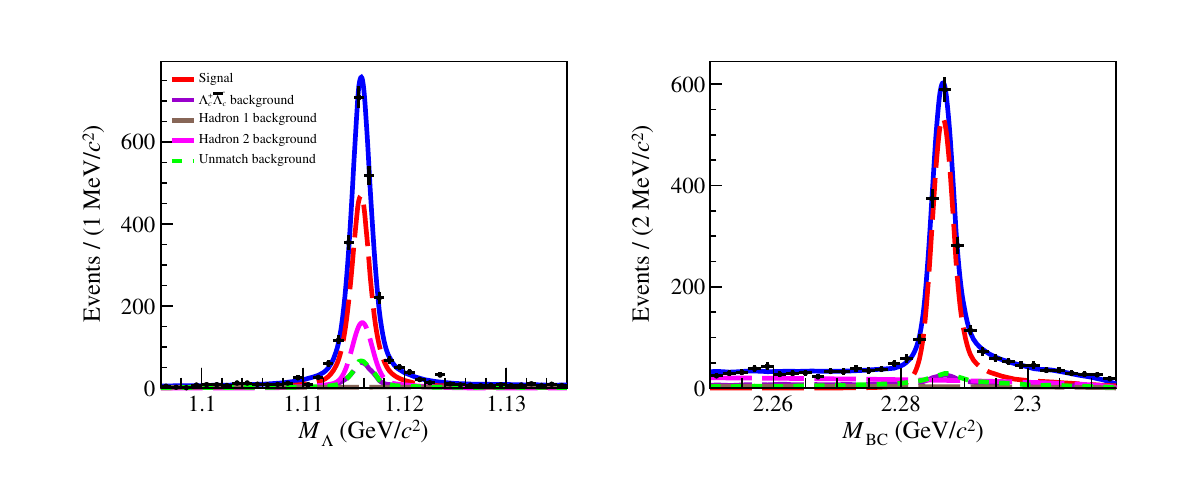}
	\vspace{-7.7mm}
	\caption{Projections of the 2D simultaneous fit on the $M_{\Lmd}$ versus $\mBC$ distributions for the data taken at $\sqrt s=4.628$ GeV. Points with error bars are data, and the colored curves show the total fit and its individual components.%Points with error bars are data, the blue solid lines are the sum of fitting functions, the red dashed lines are the $\LcptoLX$ signal, the purple dashed lines are the $\Lcp\Lcm$ background, the brown dashed lines are the ``hadron 1'' background, the magenta dashed lines are the ``hadron 2'' background and the green dashed lines are the unmatched background.
    }
	\label{fig:fit_2D_4626}
\end{figure*}

\begin{figure*}[!htb]
	\centering
	\includegraphics[width=0.8\textwidth]{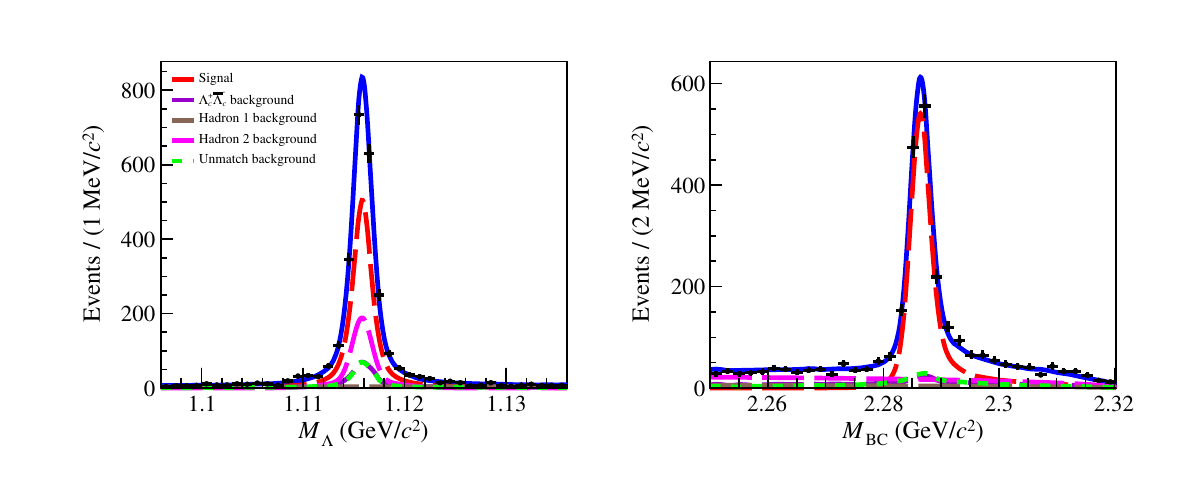}
	\vspace{-7.7mm}
	\caption{Projections of the 2D simultaneous fit on the $M_{\Lmd}$ versus $\mBC$ distributions for the data taken at $\sqrt s=4.641$ GeV. Points with error bars are data, and the colored curves show the total fit and its individual components.%Points with error bars are data, the blue solid lines are the sum of fitting functions, the red dashed lines are the $\LcptoLX$ signal, the purple dashed lines are the $\Lcp\Lcm$ background, the brown dashed lines are the ``hadron 1'' background, the magenta dashed lines are the ``hadron 2'' background and the green dashed lines are the unmatched background.
    }
	\label{fig:fit_2D_4640}
\end{figure*}

\begin{figure*}[!htb]
	\centering
	\includegraphics[width=0.8\textwidth]{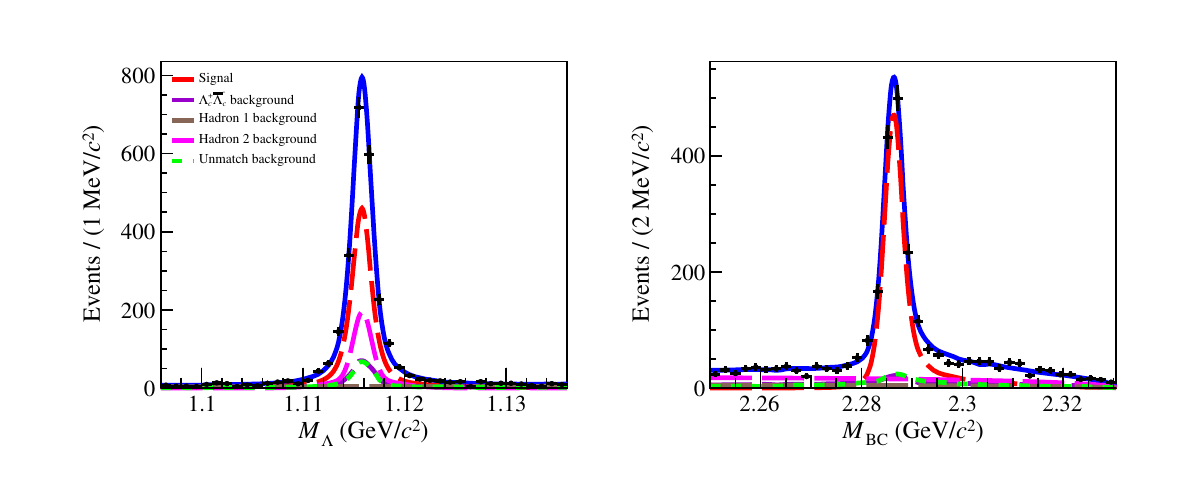}
	\vspace{-7.7mm}
	\caption{Projections of the 2D simultaneous fit on the $M_{\Lmd}$ versus $\mBC$ distributions for the data taken at $\sqrt s=4.661$ GeV. Points with error bars are data, and the colored curves show the total fit and its individual components.%Points with error bars are data, the blue solid lines are the sum of fitting functions, the red dashed lines are the $\LcptoLX$ signal, the purple dashed lines are the $\Lcp\Lcm$ background, the brown dashed lines are the ``hadron 1'' background, the magenta dashed lines are the ``hadron 2'' background and the green dashed lines are the unmatched background.
    }
	\label{fig:fit_2D_4660}
\end{figure*}

\begin{figure*}[!htb]
	\centering
	\includegraphics[width=0.8\textwidth]{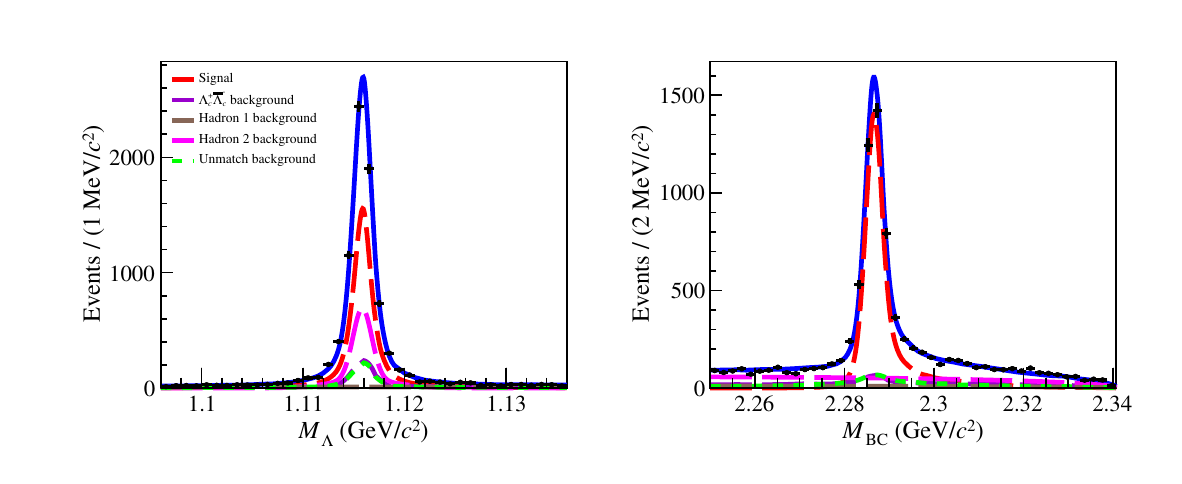}
	\vspace{-7.7mm}
	\caption{Projections of the 2D simultaneous fit on the $M_{\Lmd}$ versus $\mBC$ distributions for the data taken at $\sqrt s=4.682$ GeV. Points with error bars are data, and the colored curves show the total fit and its individual components.%Points with error bars are data, the blue solid lines are the sum of fitting functions, the red dashed lines are the $\LcptoLX$ signal, the purple dashed lines are the $\Lcp\Lcm$ background, the brown dashed lines are the ``hadron 1'' background, the magenta dashed lines are the ``hadron 2'' background and the green dashed lines are the unmatched background.
    }
	\label{fig:fit_2D_4680}
\end{figure*}

\begin{figure*}[!htb]
	\centering
	\includegraphics[width=0.8\textwidth]{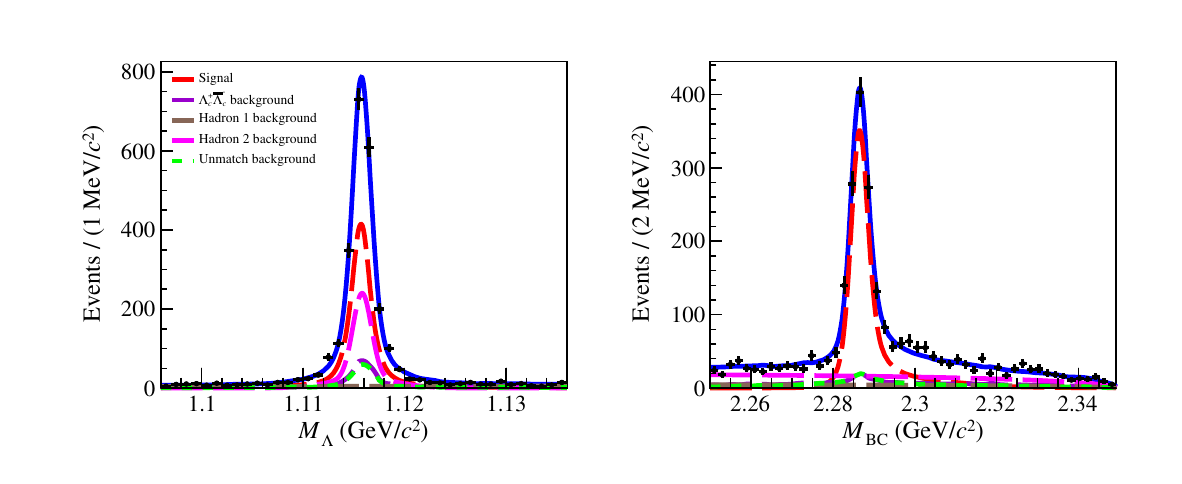}
	\vspace{-7.7mm}
	\caption{Projections of the 2D simultaneous fit on the $M_{\Lmd}$ versus $\mBC$ distributions for the data taken at $\sqrt s=4.699$ GeV. Points with error bars are data, and the colored curves show the total fit and its individual components.%Points with error bars are data, the blue solid lines are the sum of fitting functions, the red dashed lines are the $\LcptoLX$ signal, the purple dashed lines are the $\Lcp\Lcm$ background, the brown dashed lines are the ``hadron 1'' background, the magenta dashed lines are the ``hadron 2'' background and the green dashed lines are the unmatched background.
    }
	\label{fig:fit_2D_4700}
\end{figure*}

Figures \ref{fig:fit_2D_4600}, \ref{fig:fit_2D_4612}, \ref{fig:fit_2D_4626}, \ref{fig:fit_2D_4640}, \ref{fig:fit_2D_4660}, \ref{fig:fit_2D_4680} and \ref{fig:fit_2D_4700} show the projections of the 2D simultaneous fit on the $M_{\Lmd}$ versus $\mBC$ distributions in data at $\sqrt{s}=$4.600, 4.612, 4.628, 4.641, 4.661, 4.682 and 4.699 GeV, respectively.

%%%%%%%%%%%%%%%%%%%%%%%%%%%%%%%%%%%%%%%
%%%%%%%%%%%%%%%%%%%%%%%%%%%%%%%%%%%%%%%
%%%%%%%%%%%%%%%%%%%%%%%%%%%%%%%%%%%%%%%
%%%%%%%%%%%%%%%%%%%%%%%%%%%%%%%%%%%%%%%
%%%%%%%%%%%%%%%%%%%%%%%%%%%%%%%%%%%%%%%
%%%%%%%%%%%%%%%%%%%%%%%%%%%%%%%%%%%%%%%
%%%%%%%%%%%%%%%%%%%%%%%%%%%%%%%%%%%%%%%

\clearpage
\section{Results for the yields and efficiencies}
Table \ref{tab:DT_Eff} shows the DT efficiencies for each tag mode at different energy points, where the efficiencies are the overall values including both $\Lcp$ and $\Lcm$ decays.

Tables \ref{tab:Acp_lcp_STyields}, \ref{tab:Acp_lcp_eff}, and \ref{tab:DTeff_sigAm} show the ST yields, ST and DT efficiencies for each $\Lcp$ tag mode at different energy points, respectively. 

Tables \ref{tab:Acp_lcm_STyields}, \ref{tab:Acp_lcm_eff}, and \ref{tab:DTeff_sigAp} show the ST yields, ST and DT efficiencies for each $\Lcm$ tag mode at different energy points, respectively. 

\begin{table*}[htbp]
	\footnotesize
	\centering
	\caption{The combined DT efficiencies $\varepsilon_{i,j}^{\rm DT}$ for each tag mode at different energy points, where the uncertainties are statistical only.}
	\setlength\tabcolsep{7pt}
	\renewcommand{\arraystretch}{1.1}
	\begin{tabular}{
			l
			S[table-format=2.1(1)]
			S[table-format=2.1(1)]
			S[table-format=2.1(1)]
			S[table-format=2.1(1)]
			S[table-format=2.1(1)]
			S[table-format=2.1(1)]
			S[table-format=2.1(1)]
		}
		\hline
		{$\varepsilon_{i,j}^{\rm DT}(\%)$} &
		{$4.600\gev$} &
		{$4.612\gev$} &
		{$4.628\gev$} &
		{$4.641\gev$} &
		{$4.661\gev$} &
		{$4.682\gev$} &
		{$4.699\gev$} \\
		\hline
		$\Modea$   & 22.9(3) & 21.4(7) & 20.4(3) & 20.2(3) & 20.0(3) & 20.0(2) & 18.3(3) \\
		$\Modeb$   & 20.5(1) & 19.1(3) & 20.1(1) & 18.5(1) & 18.9(1) & 19.0(1) & 17.9(1) \\
		$\Modec$   & 8.4(2)  & 7.9(4)  & 7.8(2)  & 7.5(2)  & 7.6(2)  & 7.6(1)  & 7.0(2)  \\
		$\Moded$   & 8.2(2)  & 7.1(4)  & 7.2(2)  & 7.2(2)  & 7.2(2)  & 7.2(1)  & 6.8(2)  \\
		$\Modee$   & 6.5(1)  & 5.8(2)  & 6.0(1)  & 5.7(1)  & 5.9(1)  & 6.1(1)  & 5.8(1)  \\
		$\Modeaa$  & 19.4(2) & 17.8(5) & 17.1(2) & 16.3(2) & 16.7(2) & 15.5(1) & 15.2(2) \\
		$\Modebb$  & 10.6(1) & 9.8(2)  & 9.8(1)  & 9.2(1)  & 9.4(1)  & 9.3(1)  & 8.7(1)  \\
		$\Modedd$  & 7.4(1)  & 6.9(2)  & 7.0(1)  & 6.4(1)  & 6.8(1)  & 7.1(1)  & 6.5(1)  \\
		$\Modeaaa$ & 12.9(2) & 11.9(5) & 11.9(2) & 11.5(2) & 11.0(2) & 11.3(1) & 10.7(2) \\
		$\Modeccc$ & 8.6(3)  & 8.4(6)  & 8.6(3)  & 8.5(3)  & 8.1(3)  & 8.2(2)  & 7.7(3)  \\
		$\Modeddd$ & 9.0(1)  & 8.8(3)  & 8.8(1)  & 8.2(1)  & 8.2(1)  & 8.3(1)  & 7.5(1)  \\
		\hline
	\end{tabular}
	\label{tab:DT_Eff}
\end{table*}

\begin{table}[htbp]
	\centering
	\footnotesize
	\caption{The ST yields $N_{i,j}^{\rm ST}$ for each $\Lcp$ tag mode at different energy points, where the uncertainties are statistical only.}
	\setlength\tabcolsep{7pt}
	\renewcommand{\arraystretch}{1.1}
	\begin{tabular}{
			l
			S[table-format=4.0(2)]
			S[table-format=3.0(2)]
			S[table-format=4.0(2)]
			S[table-format=4.0(2)]
			S[table-format=4.0(2)]
			S[table-format=5.0(2)]
			S[table-format=4.0(2)]
		}
		\hline
		{$\Lcp\to$ tag modes} &
		{$4.600\gev$} &
		{$4.612\gev$} &
		{$4.628\gev$} &
		{$4.641\gev$} &
		{$4.661\gev$} &
		{$4.682\gev$} &
		{$4.699\gev$} \\
		\hline
		$\modea$   & 645(26)  & 113(11)  & 515(24)  & 557(25)  & 537(24)  & 1645(43) & 458(23) \\
		$\modeb$   & 3295(65) & 592(28)  & 2909(62) & 3136(64) & 3025(62) & 8572(104)& 2486(56) \\
		$\modec$   & 291(24)  & 65(12)   & 300(28)  & 288(26)  & 290(27)  & 870(46)  & 224(25) \\
		$\moded$   & 321(26)  & 46(11)   & 261(25)  & 252(24)  & 297(25)  & 760(42)  & 232(24) \\
		$\modee$   & 964(53)  & 165(23)  & 840(50)  & 744(51)  & 729(50)  & 2392(92) & 672(52) \\
		$\modeaa$  & 377(20)  & 64(8)    & 330(20)  & 360(20)  & 327(19)  & 1049(34) & 259(17) \\
		$\modebb$  & 858(40)  & 146(16)  & 750(37)  & 823(40)  & 727(36)  & 2204(63) & 636(35) \\
		$\modedd$  & 418(27)  & 80(12)   & 297(24)  & 375(26)  & 428(33)  & 1040(45) & 321(25) \\
		$\modeaaa$ & 250(18)  & 53(8)    & 171(15)  & 211(17)  & 223(17)  & 733(30)  & 175(15) \\
		$\modeccc$ & 167(18)  & 43(11)   & 149(17)  & 152(18)  & 131(17)  & 456(32)  & 120(17) \\
		$\modeddd$ & 587(34)  & 125(17)  & 438(32)  & 560(36)  & 495(34)  & 1515(62) & 479(37) \\
		\hline
	\end{tabular}
	\label{tab:Acp_lcp_STyields}
\end{table}
\begin{table*}[htbp]
	\centering
	\footnotesize
	\caption{The ST efficiencies $\varepsilon_{i,j}^{\rm ST}(\%)$ for each $\Lcp$ tag mode at different energy points, where the uncertainties are statistical only.}
	\setlength{\tabcolsep}{6pt} % 调整列间距
	\renewcommand{\arraystretch}{1.1} % 调整行高
	\begin{tabular}{
			l
			S[table-format=2.1(1)]
			S[table-format=2.1(1)]
			S[table-format=2.1(1)]
			S[table-format=2.1(1)]
			S[table-format=2.1(1)]
			S[table-format=2.1(1)]
			S[table-format=2.1(1)]
		}
		\hline
		{$\Lcp\to$ tag modes} &
		{$4.600\gev$} &
		{$4.612\gev$} &
		{$4.628\gev$} &
		{$4.641\gev$} &
		{$4.661\gev$} &
		{$4.682\gev$} &
		{$4.699\gev$} \\
		\hline
		$\modea$   & 56.1(3) & 53.4(8) & 51.8(3) & 50.7(3) & 49.7(3) & 48.6(2) & 47.6(3) \\
		$\modeb$   & 51.5(1) & 51.2(3) & 49.4(1) & 49.1(1) & 48.4(1) & 47.5(1) & 47.0(1) \\
		$\modec$   & 22.7(2) & 23.0(6) & 20.9(2) & 20.8(2) & 19.7(2) & 19.2(1) & 18.6(2) \\
		$\moded$   & 24.0(3) & 21.5(6) & 21.5(3) & 21.8(3) & 21.4(3) & 22.0(2) & 19.4(3) \\
		$\modee$   & 19.3(1) & 19.8(3) & 18.2(1) & 17.2(1) & 17.7(1) & 17.1(1) & 17.3(1) \\
		$\modeaa$  & 47.6(4) & 45.5(9) & 41.6(4) & 40.5(4) & 40.1(4) & 40.1(2) & 37.9(4) \\
		$\modebb$  & 20.8(1) & 18.9(3) & 18.5(1) & 18.6(1) & 18.4(1) & 17.6(1) & 17.5(1) \\
		$\modedd$  & 16.0(2) & 13.7(4) & 14.1(2) & 14.4(2) & 14.2(2) & 14.2(1) & 14.8(2) \\
		$\modeaaa$ & 28.0(3) & 24.5(8) & 25.8(3) & 25.2(3) & 25.4(3) & 24.7(2) & 23.4(3) \\
		$\modeccc$ & 22.8(4) & 21.5(8) & 22.4(4) & 24.9(4) & 22.4(4) & 22.2(2) & 21.4(4) \\
		$\modeddd$ & 25.1(2) & 25.2(5) & 23.2(2) & 22.8(2) & 22.9(2) & 22.3(1) & 22.1(2) \\
		\hline
	\end{tabular}
	\label{tab:Acp_lcp_eff}
\end{table*}
\begin{table*}[htbp]
	\centering
	\footnotesize
	\caption{The DT efficiencies $\varepsilon_{i,j}^{\rm DT}(\%)$ for each $\Lcp$ tag mode at different energy points, where the uncertainties are statistical only.}
	\setlength{\tabcolsep}{6pt} % 调整列间距
	\renewcommand{\arraystretch}{1.1} % 调整行高
	\begin{tabular}{
			l
			S[table-format=2.1(1)]
			S[table-format=2.1(1)]
			S[table-format=2.1(1)]
			S[table-format=2.1(1)]
			S[table-format=2.1(1)]
			S[table-format=2.1(1)]
			S[table-format=2.1(1)]
		}
		\hline
		{$\Lcp\to$ tag modes} &
		{$4.600\gev$} &
		{$4.612\gev$} &
		{$4.628\gev$} &
		{$4.641\gev$} &
		{$4.661\gev$} &
		{$4.682\gev$} &
		{$4.699\gev$} \\
		\hline
		$\modea$   & 22.2(4) & 19.7(9) & 20.5(4) & 20.1(4) & 19.5(4) & 19.7(2) & 18.0(4) \\
		$\modeb$   & 20.1(2) & 19.6(4) & 20.1(2) & 18.4(2) & 18.8(2) & 19.0(1) & 17.9(2) \\
		$\modec$   & 8.0(2)  & 8.1(5)  & 7.7(2)  & 7.4(2)  & 7.3(2)  & 7.8(1)  & 7.2(2)  \\
		$\moded$   & 7.9(3)  & 7.2(6)  & 7.6(3)  & 7.8(3)  & 7.4(3)  & 7.2(1)  & 6.8(3)  \\
		$\modee$   & 6.4(1)  & 6.2(3)  & 6.0(1)  & 5.8(1)  & 6.0(1)  & 6.0(1)  & 5.8(1)  \\
		$\modeaa$  & 18.5(3) & 17.5(7) & 16.2(3) & 15.5(3) & 15.9(3) & 14.9(2) & 14.3(3) \\
		$\modebb$  & 10.8(1) & 9.8(2)  & 9.8(1)  & 9.1(1)  & 9.3(1)  & 9.1(1)  & 8.7(1)  \\
		$\modedd$  & 7.6(1)  & 7.0(2)  & 7.1(1)  & 6.5(1)  & 7.0(1)  & 7.3(1)  & 6.5(1)  \\
		$\modeaaa$ & 12.8(3) & 9.7(6)  & 11.4(3) & 11.1(3) & 10.6(3) & 11.1(2) & 10.3(3) \\
		$\modeccc$ & 8.1(4)  & 8.6(8)  & 8.2(4)  & 8.2(4)  & 8.1(4)  & 8.2(2)  & 7.7(4)  \\
		$\modeddd$ & 9.0(2)  & 8.7(4)  & 8.8(2)  & 8.3(2)  & 8.3(2)  & 8.3(1)  & 7.4(2)  \\
		\hline
	\end{tabular}
	\label{tab:DTeff_sigAm}
\end{table*}

\begin{table*}[htbp]
	\centering
	\footnotesize
	\caption{The ST yields $N_{i,j}^{\rm ST}$ for each $\Lcm$ tag mode at different energy points, where the uncertainties are statistical only.}
	\setlength{\tabcolsep}{6pt} % 调整列间距
	\renewcommand{\arraystretch}{1.1} % 调整行高
	\begin{tabular}{
			l
			S[table-format=4.0(3)]
			S[table-format=4.0(3)]
			S[table-format=4.0(3)]
			S[table-format=4.0(3)]
			S[table-format=4.0(3)]
			S[table-format=5.0(3)]
			S[table-format=4.0(3)]
		}
		\hline
		{$\Lcm\to$ tag modes} &
		{$4.600\gev$} &
		{$4.612\gev$} &
		{$4.628\gev$} &
		{$4.641\gev$} &
		{$4.661\gev$} &
		{$4.682\gev$} &
		{$4.699\gev$} \\
		\hline
		$\Modea$   & 633(26)  & 126(12)  & 540(25)  & 552(25)  & 582(25)  & 1734(44) & 501(24) \\
		$\Modeb$   & 3516(64) & 576(27)  & 2992(62) & 3125(63) & 2924(60) & 8970(104)& 2699(57) \\
		$\Modec$   & 318(24)  & 62(11)   & 296(24)  & 315(25)  & 298(24)  & 922(43)  & 245(23) \\
		$\Moded$   & 292(23)  & 60(11)   & 235(21)  & 276(22)  & 260(22)  & 788(38)  & 234(21) \\
		$\Modee$   & 1193(55) & 192(22)  & 748(47)  & 973(52)  & 1002(53) & 2664(88) & 713(48) \\
		$\Modeaa$  & 380(20)  & 56(8)    & 346(20)  & 345(20)  & 344(20)  & 1028(34) & 280(18) \\
		$\Modebb$  & 888(39)  & 164(17)  & 730(36)  & 798(37)  & 770(36)  & 2202(61) & 685(34) \\
		$\Modedd$  & 355(24)  & 58(10)   & 291(22)  & 374(25)  & 349(24)  & 1048(42) & 330(24) \\
		$\Modeaaa$ & 276(19)  & 49(8)    & 243(16)  & 237(18)  & 233(18)  & 670(29)  & 197(16) \\
		$\Modeccc$ & 149(17)  & 31(7)    & 119(16)  & 143(17)  & 168(18)  & 432(30)  & 132(17) \\
		$\Modeddd$ & 621(39)  & 95(15)   & 561(33)  & 520(34)  & 558(34)  & 1616(60) & 480(33) \\
		\hline
	\end{tabular}
	\label{tab:Acp_lcm_STyields}
\end{table*}
\begin{table*}[htbp]
	\centering
	\footnotesize
	\caption{The ST efficiencies $\varepsilon_{i,j}^{\rm ST}(\%)$ for each $\Lcm$ tag mode at different energy points, where the uncertainties are statistical only.}
	\setlength{\tabcolsep}{6pt} % 调整列间距
	\renewcommand{\arraystretch}{1.1} % 调整行高
	\begin{tabular}{
			l
			S[table-format=2.1(1)]
			S[table-format=2.1(1)]
			S[table-format=2.1(1)]
			S[table-format=2.1(1)]
			S[table-format=2.1(1)]
			S[table-format=2.1(1)]
			S[table-format=2.1(1)]
		}
		\hline
		{$\Lcm\to$ tag modes} &
		{$4.600\gev$} &
		{$4.612\gev$} &
		{$4.628\gev$} &
		{$4.641\gev$} &
		{$4.661\gev$} &
		{$4.682\gev$} &
		{$4.699\gev$} \\
		\hline
		$\Modea$   & 56.3(3) & 54.0(8) & 51.8(3) & 50.9(3) & 49.6(3) & 48.7(2) & 47.6(3) \\
		$\Modeb$   & 51.4(1) & 51.0(3) & 49.2(1) & 48.2(1) & 48.2(1) & 46.8(1) & 45.7(1) \\
		$\Modec$   & 23.3(2) & 21.6(6) & 20.8(2) & 20.9(2) & 20.7(2) & 20.4(1) & 19.5(2) \\
		$\Moded$   & 23.1(3) & 22.2(6) & 19.9(3) & 20.1(3) & 20.8(3) & 19.6(2) & 19.8(3) \\
		$\Modee$   & 22.2(1) & 20.7(3) & 19.5(1) & 19.6(1) & 19.1(1) & 18.9(1) & 18.2(1) \\
		$\Modeaa$  & 49.2(4) & 48.4(9) & 44.6(4) & 45.2(4) & 43.3(4) & 42.6(2) & 40.8(4) \\
		$\Modebb$  & 21.8(1) & 20.7(3) & 19.8(1) & 19.6(1) & 19.4(1) & 18.8(1) & 18.5(1) \\
		$\Modedd$  & 15.3(2) & 13.5(4) & 13.7(2) & 14.1(2) & 14.0(2) & 13.8(1) & 14.5(2) \\
		$\Modeaaa$ & 30.9(4) & 28.9(8) & 28.7(4) & 27.1(3) & 27.6(4) & 27.2(2) & 25.3(4) \\
		$\Modeccc$ & 24.5(4) & 23.6(9) & 24.5(4) & 24.8(4) & 24.0(4) & 23.1(2) & 23.0(4) \\
		$\Modeddd$ & 25.8(2) & 26.3(5) & 23.7(2) & 23.9(2) & 23.4(2) & 22.4(1) & 22.9(2) \\
		\hline
	\end{tabular}
	\label{tab:Acp_lcm_eff}
\end{table*}
\begin{table*}[htbp]
	\centering
	\footnotesize
	\caption{The DT efficiencies $\varepsilon_{i,j}^{\rm DT}(\%)$ for each $\Lcm$ tag mode at different energy points, where the uncertainties are statistical only.}
	\setlength{\tabcolsep}{6pt} % 调整列间距
	\renewcommand{\arraystretch}{1.1} % 调整行高
	\begin{tabular}{
			l
			S[table-format=2.1(1)]
			S[table-format=2.1(1)]
			S[table-format=2.1(1)]
			S[table-format=2.1(1)]
			S[table-format=2.1(1)]
			S[table-format=2.1(1)]
			S[table-format=2.1(1)]
		}
		\hline
		{$\Lcm\to$ tag modes} &
		{$4.600\gev$} &
		{$4.612\gev$} &
		{$4.628\gev$} &
		{$4.641\gev$} &
		{$4.661\gev$} &
		{$4.682\gev$} &
		{$4.699\gev$} \\
		\hline
		$\Modea$   & 23.7(4) & 22.9(9) & 20.3(4) & 20.4(4) & 20.4(4) & 20.2(2) & 18.7(4) \\
		$\Modeb$   & 21.1(2) & 18.7(4) & 20.2(2) & 18.6(2) & 19.1(2) & 19.1(1) & 18.0(2) \\
		$\Modec$   & 8.8(3)  & 7.7(5)  & 7.9(3)  & 7.7(2)  & 7.9(3)  & 7.4(1)  & 6.8(2)  \\
		$\Moded$   & 8.4(3)  & 7.1(6)  & 6.7(3)  & 6.5(2)  & 6.9(3)  & 7.2(2)  & 6.8(3)  \\
		$\Modee$   & 6.6(1)  & 5.4(3)  & 6.0(1)  & 5.6(1)  & 5.7(1)  & 6.1(1)  & 5.7(1)  \\
		$\Modeaa$  & 20.4(4) & 18.0(7) & 18.0(3) & 17.1(3) & 17.6(3) & 16.1(2) & 16.1(3) \\
		$\Modebb$  & 10.4(1) & 9.9(2)  & 10.0(1) & 9.3(1)  & 9.5(1)  & 9.4(1)  & 8.7(1)  \\
		$\Modedd$  & 7.2(1)  & 6.7(3)  & 6.8(1)  & 6.3(1)  & 6.5(1)  & 6.9(1)  & 6.5(1)  \\
		$\Modeaaa$ & 13.1(3) & 14.2(7) & 12.5(3) & 11.9(3) & 11.4(3) & 11.6(2) & 11.2(3) \\
		$\Modeccc$ & 9.2(4)  & 8.2(8)  & 9.0(4)  & 8.8(4)  & 8.1(4)  & 8.2(2)  & 7.8(4)  \\
		$\Modeddd$ & 9.1(2)  & 8.9(4)  & 8.9(2)  & 8.1(2)  & 8.2(2)  & 8.3(1)  & 7.7(2)  \\
		\hline
	\end{tabular}
	\label{tab:DTeff_sigAp}
\end{table*}
%%%%%%%%%%%%%%%%%%%%%%%%%%%%%%%%%%%%%%%%%%%%%%%%%%%%%%%%%

\clearpage
\section{One-dimensional efficiency projections and corrected angular distributions}

\begin{figure*}[!htb]
	\centering
	\includegraphics[width=0.37\textwidth]{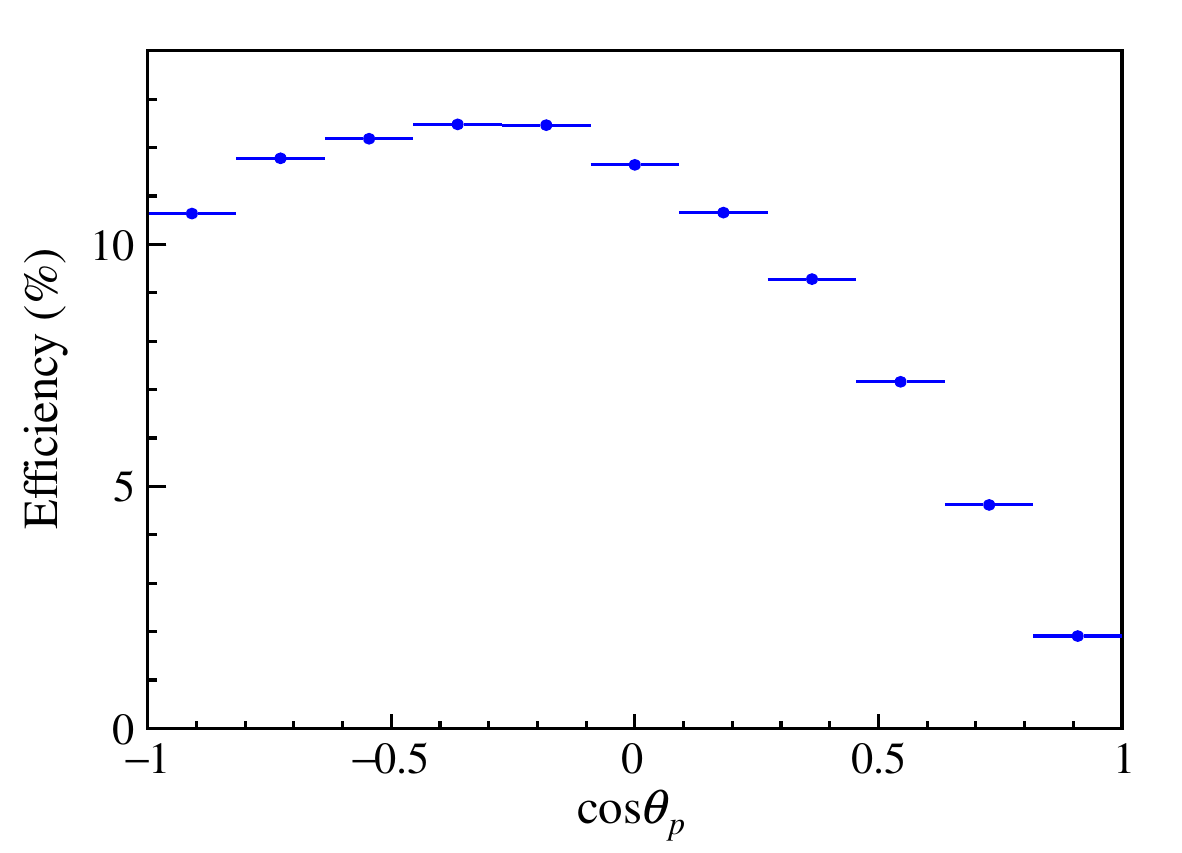}\hspace{0.020\textwidth}
	\includegraphics[width=0.37\textwidth]{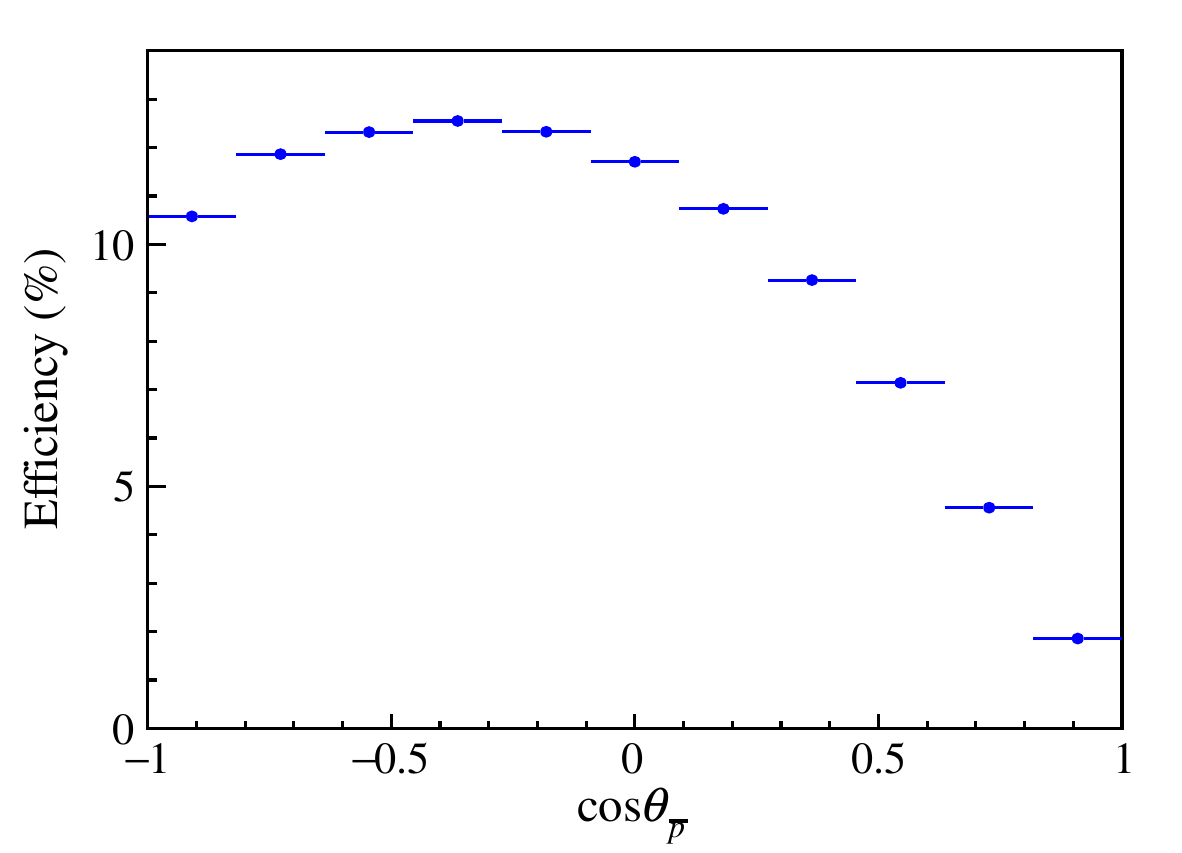}
	\vspace{-5.0mm}
	\caption{One-dimensional projections of the detection efficiencies as functions
of (left) $\cos\theta_p$ for $\Lambda_c^+\to\Lambda X$ and
(right) $\cos\theta_{\bar p}$ for
$\bar{\Lambda}_c^-\to\bar{\Lambda}X$.}
    \label{fig:mc_eff}
\end{figure*}

\begin{figure*}[!htb]
	\centering
	\includegraphics[width=0.4\textwidth]{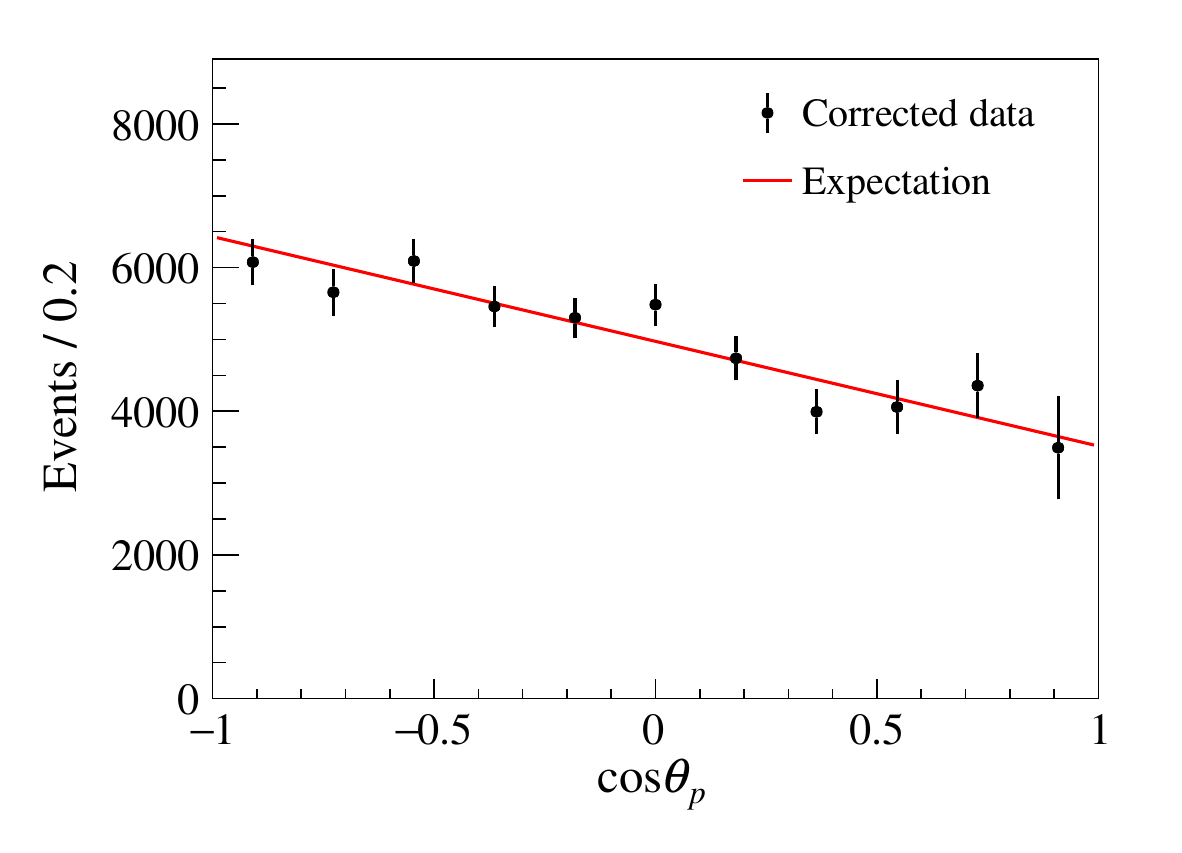}
	\includegraphics[width=0.4\textwidth]{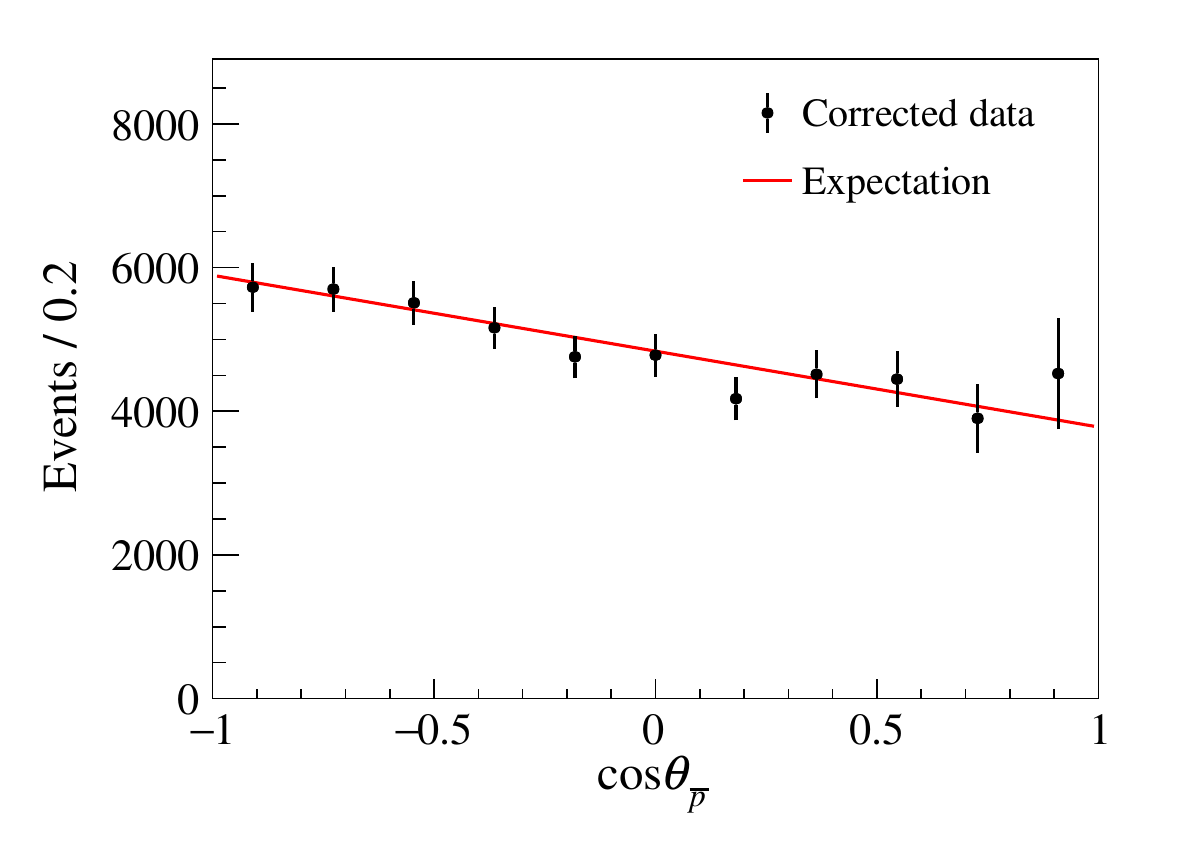}
	\vspace{-5.7mm}
	\caption{Background-subtracted and efficiency-corrected distributions of   $\cos\theta_{p}$ for (left) $\Lambda_c^+ \to \Lambda X$ and (right) $\cos\theta_{\bar{p}}$ for $\bar{\Lambda}_c^- \to \bar{\Lambda} X$. The points with error bars are the corrected data, and the red lines show the expectation calculated from likelihood fit results.}
    \label{fig:data_eff_correct}
\end{figure*}

The one-dimensional projections of the detection efficiencies as
functions of $\cos\theta_p$ and $\cos\theta_{\bar{p}}$ are shown in
Fig.~\ref{fig:mc_eff}.
Figure~\ref{fig:data_eff_correct} presents the background-subtracted and
efficiency-corrected $\cos\theta_p$ and $\cos\theta_{\bar{p}}$ distributions, together with the
expectations obtained from the unbinned maximum-likelihood fit. The
agreement between the corrected data and the fit expectations indicates
that the extracted polarization provides a consistent description of the
observed angular distributions.


%% file: draft_Lc2LmdX_arXiv.bbl
\begin{thebibliography}{99}

\bibitem{Sakharov:1991}
A.~D.~Sakharov,
%%\emph{Violation of CP Invariance, C asymmetry, and baryon asymmetry of the universe},
\href{https://doi.org/10.1070/PU1991v034n05ABEH002497}{Pisma Zh. Eksp. Teor. Fiz. \textbf{5}, 32 (1967)}.

%\cite{Christenson:1964fg}
\bibitem{ref:KsCPV}
J.~H.~Christenson, J.~W.~Cronin, V.~L.~Fitch and R.~Turlay,
%%\emph{Evidence for the $2\pi$ Decay of the $K_2^0$ Meson},
\href{https://doi.org/10.1103/PhysRevLett.13.138}{Phys. Rev. Lett. \textbf{13}, 138 (1964).}

\bibitem{ref:BCPV_BaBar}
B.~Aubert \textit{et al.} (BaBar Collaboration),
%%\emph{Observation of CP violation in the $B^0$ meson system},
\href{https://doi.org/10.1103/PhysRevLett.87.091801}{Phys. Rev. Lett. \textbf{87}, 091801 (2001).}

%\cite{Belle:2001zzw}
\bibitem{ref:BCPV_Belle}
K.~Abe \textit{et al.} (Belle Collaboration),
%%\emph{Observation of large CP violation in the neutral $B$ meson system},
\href{https://doi.org/10.1103/PhysRevLett.87.091802}{Phys. Rev. Lett. \textbf{87}, 091802 (2001).}

%\cite{LHCb:2019hro}
\bibitem{LHCb:2019hro}
R.~Aaij \textit{et al.} (LHCb Collaboration),
%%\emph{Observation of CP Violation in Charm Decays},
\href{https://doi.org/10.1103/PhysRevLett.122.211803}{Phys. Rev. Lett. \textbf{122}, 211803 (2019).}

%\cite{LHCb:2025ray}
\bibitem{LHCb:LbCPV}
R.~Aaij \textit{et al.} (LHCb Collaboration),
%``Observation of charge{\textendash}parity symmetry breaking in baryon decays,''
\href{https://doi.org/10.1038/s41586-025-09119-3}{Nature \textbf{643}, 1223 (2025).}

%%%%%%%%%%%%%%%%%%%%%%%%%%%%%%%
%\cite{Cheng:2012wr}
\bibitem{Cheng:2012wr}
H.~Y.~Cheng and C.~W.~Chiang,
%``Direct CP violation in two-body hadronic charmed meson decays,''
\href{https://doi.org/10.1103/PhysRevD.85.034036}{Phys. Rev. D \textbf{85}, 034036 (2012).}
%[erratum: Phys. Rev. D \textbf{85}, 079903 (2012)]


%\cite{Cheng:2021qpd}
\bibitem{Cheng:2021qpd}
H.~Y.~Cheng,
%``Charmed baryon physics circa 2021,''
\href{https://doi.org/10.1016/j.cjph.2022.06.021}{Chin. J. Phys. \textbf{78}, 324 (2022).}
%
%\cite{Li:2025nzx}
\bibitem{Li:2025nzx}
P.~R.~Li, X.~R.~Lyu and Y.~H.~Zheng,
%``Experimental overview on the charmed baryon decays*,''
\href{https://doi.org/10.1088/1674-1137/ae1187}{Chin. Phys. C \textbf{50}, 022002 (2026).}
%
%%\cite{Delepine:2019cpp}
%\bibitem{Delepine:2019cpp}
%D.~Delepine, G.~Faisel and C.~A.~Ramirez,
%%``Direct CP violation in $D^+ \rightarrow K^0({\bar{K}}^0) \pi ^+$ decays as a probe for new physics,''
%\href{https://doi.org/10.1140/epjc/s10052-020-8150-0}{Eur. Phys. J. C \textbf{80}, no.7, 596 (2020).}
%
%%\cite{Dery:2019ysp}
%\bibitem{Dery:2019ysp}
%A.~Dery and Y.~Nir,
%%``Implications of the LHCb discovery of CP violation in charm decays,''
%\href{https://doi.org/10.1007/JHEP12(2019)104}{JHEP \textbf{12}, 104 (2019).}
%
%%\cite{Chala:2019fdb}
%\bibitem{Chala:2019fdb}
%M.~Chala, A.~Lenz, A.~V.~Rusov and J.~Scholtz,
%%``$\Delta A_{CP}$ within the Standard Model and beyond,''
%\href{https://doi.org/10.1007/JHEP07(2019)161}{JHEP \textbf{07}, 161 (2019).}
%%92 citations counted in INSPIRE as of 01 Aug 2025
%
\bibitem{Capstick:2000qj}
S.~Capstick and W.~Roberts,
%``Quark models of baryon masses and decays,''
\href{https://doi.org/10.1016/S0146-6410(00)00109-5}{Prog. Part. Nucl. Phys. \textbf{45}, S241 (2000).}

%\bibitem{pdg:2024cfk}
%S.~Navas \textit{et al.} (Particle Data Group),
%%\emph{Review of particle physics},
%\href{https://doi.org/10.1103/PhysRevD.110.030001}{Phys. Rev. D \textbf{110}, 030001 (2024) and 2025 update.}

\bibitem{pdg:2024cfk}
F. Takahashi \textit{et al.} (Particle Data Group), Int. J. Mod. Phys. A 41, 2630011 (2026).

%\cite{BESIII:2018ciw}
\bibitem{BESIII:2018Lmdx}
M.~Ablikim \textit{et al.} (BESIII Collaboration),
%\emph{Measurement of absolute branching fraction of the inclusive decay $\Lambda_{c}^{+} \to \Lambda + X$},
\href{https://doi.org/10.1103/PhysRevLett.121.062003}{Phys. Rev. Lett. \textbf{121}, 062003 (2018).}

%%\cite{Falk:1992ws}
%\bibitem{Falk:1992ws}
%A.~F.~Falk and M.~Neubert,
%%``Second order power corrections in the heavy quark effective theory. 2. Baryon form-factors,''
%\href{https://doi.org/10.1103/PhysRevD.47.2982}{Phys. Rev. D \textbf{47}, 2982 (1993).}
%%[arXiv:hep-ph/9209269 [hep-ph]].
%146 citations counted in INSPIRE as of 10 May 2025

\bibitem{Wang:2024wrm}
H.~J.~Wang, P.~R.~Li, X.~R.~Lyu, J.~Tandean and H.~B.~Li,
\href{https://doi.org/10.1016/j.scib.2025.02.030}{Sci. Bull. \textbf{70}, 1183 (2025).}
%%%%%%%%%%%%%%%%%%%%%%%%%%%%%%%%

%\cite{LHCb:2024tnq}
\bibitem{LHCb:Lmdpi_alpha}
R.~Aaij \textit{et al.} (LHCb Collaboration),
%%\emph{Measurement of \ensuremath{\Lambda}b0, \ensuremath{\Lambda}c+, and \ensuremath{\Lambda} Decay Parameters Using \ensuremath{\Lambda}b0\textrightarrow{}\ensuremath{\Lambda}c+h- Decays},
\href{https://doi.org/10.1103/PhysRevLett.133.261804}{Phys. Rev. Lett. \textbf{133}, 261804 (2024).}

%\cite{Belle:2022uod}
\bibitem{Belle:LmdSgmh}
L.~K.~Li \textit{et al.} (Belle Collaboration),
%%\emph{Search for CP violation and measurement of branching fractions and decay asymmetry parameters for $\Lcp\to\Lmd h^{+}$ and $\Lcp\to\Sigma^0h^+ (h=K,\pi)$},
\href{https://doi.org/10.1016/j.scib.2023.02.017}{Sci. Bull. \textbf{68}, 583 (2023).}

%\cite{Wang:2022tcm}
\bibitem{Wang:2022tcm}
J.~P.~Wang and F.~S.~Yu,
%%\emph{Probing hyperon CP violation with charmed baryon decays},
\href{https://doi.org/10.1016/j.physletb.2024.138460}{Phys. Lett. B \textbf{849}, 138460 (2024).}

\bibitem{ref:DT_method}
J.~Adler \textit{et al.} (MARKIII Collaboration),
%\emph{Measurement of the Branching Fractions for D0 ---\ensuremath{>} pi- e+ electron-neutrino and D0 ---\ensuremath{>} K- e+ electron-neutrino and Determination of (V (c d) / V (c s))**2},
\href{https://doi.org/10.1103/PhysRevLett.62.1821}{Phys. Rev. Lett. \textbf{62}, 1821 (1989).}

\bibitem{BESIII:energy2}
M.~Ablikim \textit{et al.} (BESIII Collaboration),
%\emph{Measurement of integrated luminosities at BESIII for data samples at center-of-mass energies between 4.0 and 4.6 GeV},
\href{https://doi.org/doi:10.1088/1674-1137/ac80b4}{Chin. Phys. C \textbf{46}, 113002 (2022).}

\bibitem{BESIII:energy3}
M.~Ablikim \textit{et al.} (BESIII Collaboration),
%\emph{Luminosities and energies of $e^+e^-$ collision data taken between $\sqrt{s}$=4.612 GeV and 4.946 GeV at BESIII},
\href{https://doi.org/10.1088/1674-1137/ac84cc}{Chin. Phys. C \textbf{46}, 113003 (2022).}
%doi:10.1088/1674-1137/ac84cc
%[arXiv:2205.04809 [hep-ex]].
%16 citations counted in INSPIRE as of 08 Nov 2022

\bibitem{BESIII:2009fln}
{M.~Ablikim \textit{et al.} (BESIII Collaboration),}
%\emph{Design and Construction of the BESIII Detector},
\href{https://doi.org/10.1016/j.nima.2009.12.050}
{Nucl. Instrum. Meth. A
\textbf{614}, 345 (2010).}
%doi:10.1016/j.nima.2009.12.050
%[arXiv:0911.4960 [physics.ins-det]].
%1102 citations counted in INSPIRE as of 16 Jan 2024

\bibitem{ref:14}
C. Yu {\it et al}., in \textit{Proceedings of IPAC2016, Busan, Korea} (JACoW, Geneva, Switzerland, 2016), p. TUYA01.

\bibitem{BESIII:2020nme}
{M.~Ablikim \textit{et al.} (BESIII Collaboration),}
%\emph{Future Physics Programme of BESIII},
\href{https://doi.org/10.1088/1674-1137/44/4/040001}
{Chin. Phys. C \textbf{44}, 040001 (2020).}
%9doi:10.1088/1674-1137/44/4/040001
%[arXiv:1912.05983 [hep-ex]].
%440 citations counted in INSPIRE as of 16 Jan 2024


%\cite{Li:2017jpg}
\bibitem{Li:2017jpg}
X.~Li \textit{et al.,}
%X.~Li, Y.~Sun, C.~Li, Z.~Liu, Y.~Heng, M.~Shao, X.~Wang, Z.~Wu, P.~Cao and M.~Chen, \textit{et al.}
%\emph{Study of MRPC technology for BESIII endcap-TOF upgrade},
\href{https://doi.org/10.1007/s41605-017-0014-2}{Radiat. Detect. Technol. Methods \textbf{1}, 13 (2017)}.
%doi:10.1007/s41605-017-0014-2
%147 citations counted in INSPIRE as of 16 Jan 2024

%\cite{Guo:2017sjt}
\bibitem{Guo:2017sjt}
{Y.~X.~Guo \textit{et al.,}}
%Y.~X.~Guo, S.~S.~Sun, F.~F.~An, R.~X.~Yang, M.~Zhou, Z.~Wu, H.~L.~Dai, Y.~K.~Heng, C.~Li and Z.~Y.~Deng, \textit{et al.}
%\emph{The study of time calibration for upgraded end cap TOF of BESIII},
\href{https://doi.org/10.1007/s41605-017-0012-4}
{Radiat. Detect. Technol. Methods \textbf{1}, 15 (2017).}
%doi:10.1007/s41605-017-0012-4
%142 citations counted in INSPIRE as of 16 Jan 2024

%\cite{Cao:2020ibk}
\bibitem{Cao:2020ibk}
{P.~Cao \textit{et al.,}}
%P.~Cao, H.~F.~Chen, M.~M.~Chen, H.~L.~Dai, Y.~K.~Heng, X.~L.~Ji, X.~S.~Jiang, C.~Li, X.~Li and S.~B.~Liu, \textit{et al.}
%\emph{Design and construction of the new BESIII endcap Time-of-Flight system with MRPC Technology},
\href{https://doi.org/10.1016/j.nima.2019.163053}
{Nucl. Instrum. Meth. A \textbf{953}, 163053 (2020).}
%doi:10.1016/j.nima.2019.163053
%103 citations counted in INSPIRE as of 16 Jan 2024

%\cite{Jadach:2000ir}
\bibitem{ref:Jadach2000ir}
{S.~Jadach, B.~F.~L.~Ward, and Z.~Was,}
%\emph{Coherent exclusive exponentiation for precision Monte Carlo calculations},
\href{https://doi.org/10.1103/PhysRevD.63.113009}
{Phys. Rev. D \textbf{63}, 113009 (2001).}
%doi:10.1103/PhysRevD.63.113009
%[arXiv:hep-ph/0006359 [hep-ph]].
%721 citations counted in INSPIRE as of 16 Jan 2024


%\cite{Lange:2001uf}
\bibitem{ref:Lange2001uf}
{D.~J.~Lange,}
%\emph{The EvtGen particle decay simulation package},
\href{https://doi.org/10.1016/S0168-9002(01)00089-4}
{Nucl. Instrum. Meth. A \textbf{462}, 152 (2001).}
%doi:10.1016/S0168-9002(01)00089-4
%4081 citations counted in INSPIRE as of 16 Jan 2024

%\cite{Ping:2008zz}
\bibitem{ref:Ping2008zz}
{R.~G.~Ping,}
%\emph{Event generators at BESIII},
\href{https://doi.org/10.1088/1674-1137/32/8/001}
{Chin. Phys. C \textbf{32}, 599 (2008).}
%doi:10.1088/1674-1137/32/8/001
%442 citations counted in INSPIRE as of 16 Jan 2024


%\cite{Chen:2000tv}
\bibitem{Chen:2000tv}
{J.~C.~Chen, G.~S.~Huang, X.~R.~Qi, D.~H.~Zhang, and Y.~S.~Zhu,}
%\emph{Event generator for J / psi and psi (2S) decay},
\href{https://doi.org/10.1103/PhysRevD.62.034003}
{Phys. Rev. D \textbf{62}, 034003 (2000);} R.~L.~Yang, R.~G.~Ping and H.~Chen,
\href{https://doi.org/10.1088/0256-307X/31/6/061301}{Chin. Phys. Lett. \textbf{31}, 061301 (2014).}
%341 citations counted in INSPIRE as of 05 Oct 2025

%\cite{GEANT4:2002zbu}
\bibitem{GEANT4:2002zbu}
{S.~Agostinelli \textit{et al.} (GEANT4 Collaboration),}
%\emph{GEANT4--a simulation toolkit},
\href{https://doi.org/10.1016/S0168-9002(03)01368-8}
{Nucl. Instrum. Meth. A \textbf{506}, 250 (2003).}
%doi:10.1016/S0168-9002(03)01368-8
%18007 citations counted in INSPIRE as of 16 Jan 2024

%\cite{Allison:2006ve}
\bibitem{Allison:2006ve}
{J.~Allison \textit{et al.,}}
%J.~Allison, K.~Amako, J.~Apostolakis, H.~Araujo, P.~A.~Dubois, M.~Asai, G.~Barrand, R.~Capra, S.~Chauvie and R.~Chytracek, \textit{et al.}
%\emph{Geant4 developments and applications},
\href{https://doi.org/10.1109/TNS.2006.869826}
{IEEE Trans. Nucl. Sci. \textbf{53}, 270 (2006).}
%doi:10.1109/TNS.2006.869826
%3873 citations counted in INSPIRE as of 16 Jan 2024

%\cite{BESIII:2023rwv}
\bibitem{BESIII:2023rwv}
{M.~Ablikim \textit{et al.} (BESIII Collaboration),}
%\emph{Measurement of Energy-Dependent Pair-Production Cross Section and Electromagnetic Form Factors of a Charmed Baryon}
\href{https://doi.org/10.1103/PhysRevLett.131.191901}
{Phys. Rev. Lett. \textbf{131}, 191901 (2023).}
%doi:10.1103/PhysRevLett.131.191901
%[arXiv:2307.07316 [hep-ex]].
%17 citations counted in INSPIRE as of 04 Jul 2024

%\cite{BESIII:2022udq}
\bibitem{BESIII:2022udq}
M.~Ablikim \textit{et al.} (BESIII Collaboration),
%``Partial wave analysis of the charmed baryon hadronic decay $ {\Lambda}_c^{+} ${\textrightarrow} {\ensuremath{\Lambda}}{\ensuremath{\pi}}$^{+}${\ensuremath{\pi}}$^{0}$,''
\href{https://doi.org/10.1007/JHEP12(2022)033}{JHEP \textbf{12}, 033 (2022).}

%\cite{BESIII:2025zbz}
\bibitem{BESIII:2025zbz}
M.~Ablikim \textit{et al.} (BESIII Collaboration),
%``The Production and Decay Dynamics of the Charmed Baryon $\Lambda_c^+$ in $e^+e^-$ Annihilations near Threshold,''
\href{https://arxiv.org/abs/2508.11400}{arXiv:2508.11400.}
%1 citations counted in INSPIRE as of 29 Sep 2025

\bibitem{BESIII:xikalpha}
M.~Ablikim \textit{et al.} (BESIII Collaboration),
%\emph{First Measurement of the Decay Asymmetry in the Pure W-Boson-Exchange Decay $\Lcp\to\Xi^{0}K^{+}$ }
\href{https://doi.org/10.1103/PhysRevLett.132.031801}{Phys. Rev. Lett. \textbf{132}, 031801 (2024).}

%\cite{Rogozhnikov:2016bdp}
\bibitem{hepml}
A.~Rogozhnikov,
%``Reweighting with Boosted Decision Trees,''
\href{https://doi.org/10.1088/1742-6596/762/1/012036}{J. Phys. Conf. Ser. \textbf{762}, 012036 (2016).}

%\cite{BESIII:2022xne}
\bibitem{BESIII:nhadron}
{M.~Ablikim \textit{et al.} (BESIII Collaboration),}
%\emph{Observations of the Cabibbo-Suppressed decays $\Lambda_{c}^{+}\to n\pi^{+}\pi^{0}$, $n\pi^{+}\pi^{-}\pi^{+}$ and the Cabibbo-Favored decay $\Lambda_{c}^{+}\to nK^{-}\pi^{+}\pi^{+}$},
\href{https://doi.org/10.1088/1674-1137/ac9d29}
{Chin. Phys. C \textbf{47}, 023001 (2023).}
%doi:10.1088/1674-1137/ac9d29
%[arXiv:2210.03375 [hep-ex]].
%9 citations counted in INSPIRE as of 24 Jan 2024

%\cite{Cranmer:2000du}
\bibitem{Cranmer:2000du}
K.~S.~Cranmer,
%``Kernel estimation in high-energy physics,''
\href{https://doi.org/10.1016/S0010-4655(00)00243-5}{Comput. Phys. Commun. \textbf{136}, 198 (2001).}

%\bibitem{ARGUS:1990hfq}
%H.~Albrecht \textit{et al.} (ARGUS Collaboration),
%%\emph{Search for Hadronic $b \to u$ Decays},
%\href{https://doi.org/10.1016/0370-2693(90)91293-K}{Phys. Lett. B \textbf{241}, 278 (1990).}

\bibitem{supp}
See Supplemental Material at
\href{https://link.aps.org/supplemental/xxxx}
{https://link.aps.org/supplemental/xxxx}
for additional fit plots, ST yields and efficiencies, DT efficiencies,
and efficiency projections and corrected angular distributions.

\bibitem{Langenbruch:2019nwe}
C.~Langenbruch,
%``Parameter uncertainties in weighted unbinned maximum likelihood fits,''
\href{https://doi.org/10.1140/epjc/s10052-022-10254-8}{Eur. Phys. J. C \textbf{82}, 393 (2022).}

%\cite{James:1975dr}
\bibitem{James:1975dr}
F.~James and M.~Roos,
%\emph{Minuit: A System for Function Minimization and Analysis of the Parameter Errors and Correlations},
\href{https://doi.org/10.1016/0010-4655(75)90039-9}{Comput. Phys. Commun. \textbf{10}, 343 (1975).}


\end{thebibliography}
